\documentclass{article}
\usepackage[T1]{fontenc} 
\usepackage[utf8]{inputenc} 
\usepackage{ismir,amsmath,cite,url}
\usepackage{graphicx}
\usepackage{color}
\usepackage{subcaption}
\usepackage{paralist}
\usepackage{multicol}
\usepackage{microtype}
\usepackage{flushend}
\usepackage{amsfonts}
\usepackage{multirow}
\usepackage{booktabs}
\usepackage{tabularx}

\usepackage{lineno}

\title{Is Disentanglement enough? On Latent Representations for Controllable Music Generation}

\twoauthors
 {Ashis Pati} {Center for Music Technology \\ Georgia Institute of Technology, USA \\ {\tt \small ashis.pati@gatech.edu}} 
 {Alexander Lerch} {Center for Music Technology \\ Georgia Institute of Technology, USA \\ {\tt \small alexander.lerch@gatech.edu}}

\begin{document}

\maketitle
\begin{abstract}
  Improving \textit{controllability} or the ability to manipulate one or more attributes of the generated data has become a topic of interest in the context of deep generative models of music. 
  Recent attempts in this direction have relied on learning disentangled representations from data such that the underlying factors of variation are well separated. 
  In this paper, we focus on the relationship between disentanglement and controllability by conducting a systematic study using different supervised disentanglement learning algorithms based on the Variational Auto-Encoder (VAE) architecture. Our experiments show that a high degree of disentanglement can be achieved by using different forms of supervision to train a strong discriminative encoder. However, in the absence of a strong generative decoder, disentanglement does not necessarily imply controllability. 
  The structure of the latent space with respect to the VAE-decoder plays an important role in boosting the ability of a generative model to manipulate different attributes. 
  To this end, we also propose methods and metrics to help evaluate the quality of a latent space with respect to the afforded degree of controllability.

\end{abstract}

\section{Introduction} \label{sec:intro}
  Automatic music generation using machine learning has seen significant improvements over the last decade. Deep generative models relying on neural networks have been successfully applied to several different music generation tasks, e.g., monophonic music generation consisting of a single melodic line~\cite{roberts_hierarchical_2018,colombo2016algorithmic,sturm2016music}, polyphonic music generation involving several different parts or instruments \cite{yang2017midinet,boulanger2012modeling}, and creating musical renditions with expressive timing and dynamics \cite{huang2019musictransformer,oore2018time}. However, such models are usually found lacking in two critical aspects: \textit{controllability} and \textit{interactivity} \cite{Briot2018}.  Most of the models typically work as black-boxes, i.e., the intended end-user has little to no control over the generation process. Additionally, they do not allow any modes for interaction, i.e., the user cannot selectively modify the generated music or some of its parts based on desired musical characteristics. Consequently, there have been considerable efforts focusing on controllable music generation \cite{pati_learning_2019, yang2019deep, huang2019counterpoint} in interactive settings \cite{hadjeres2017deepbach,donahue2019piano,bazin2019nonoto}. One promising avenue for enabling controllable music generation stems from the field of representation learning. 

  Representation learning involves automatic extraction of the underlying factors of variation in given data \cite{bengio_representation_2013}. The majority of the current state-of-the-art machine learning-based methods aim at learning compact and useful representations \cite{chen2020simple, caron2020unsupervised}. These have been used for solving different types of discriminative or generative tasks spanning several domains such as images, text, speech, audio, and music. A  special case of representation learning deals with \textit{disentangled} representations, where individual factors of variation are clearly separated such that changes to a single underlying factor in the data lead to changes in a single factor of the learned disentangled representation \cite{locatello_challenging_2019}. Specifically, in the context of music, disentangled representations have been used for a wide variety of music generation tasks such as rhythm transfer \cite{yang2019deep, jiang2020transformer}, genre transfer \cite{brunner_midi-vae_2018}, instrument rearrangement \cite{hung2019musical}, timbre synthesis \cite{luo2019learning}, and manipulating low-level musical attributes \cite{hadjeres_glsr-vae_2017, pati19latent-reg, tan2020music}. 
  
  Disentangled representation learning has been an active area of research in the context of deep generative models for music. Previous methods have focused on different types of musical attributes (e.g., note density \cite{hadjeres_glsr-vae_2017}, rhythm \cite{yang2019deep}, timbre \cite{luo2019learning}, genre \cite{brunner_midi-vae_2018}, and arousal \cite{tan2020music}) and have achieved promising results. However, contrary to other fields such as computer vision \cite{locatello_challenging_2019, Locatello2020Disentangling}, research on disentanglement learning in the context of music has been task-specific and ad-hoc. Consequently, the degree to which disentangled representations can aid controllable music generation remains largely unexplored. While we have shown that unsupervised disentanglement learning methods are not suitable for music-based tasks \cite{pati2020dmelodies}, the use of supervised learning methods has not been systematically evaluated.   

  In this paper, we conduct a systematic study on controllable generation by using supervised methods to learn disentangled representations. We compare the performance of several supervised methods and conduct a series of experiments to objectively evaluate their performance in terms of disentanglement and controllability for music generation. 
  In the context of this paper, \textit{controllability} is defined as the ability of a generative model to selectively, independently, and predictably manipulate one or more attributes (for instance, rhythm, scale) of the generated data. We show that while supervised learning methods can achieve a high degree of disentanglement in the learned representation, not all methods are equally useful from the perspective of controllable generation. 
  The degree of controllability depends not only on the learning methods but also on the musical attribute to be controlled.  
  In order to foster reproducibility, the code for the conducted experiments is available online.\footnote{https://github.com/ashispati/dmelodies\_controllability \\ last accessed: 1st Aug 2021 \label{footnote1}} 

\section{Method \& Experimental Setup} \label{sec:method}

  The primary goal of this paper is to investigate the degree to which learning disentangled representations can provide control over manipulating different attributes of the generated music. To this end, we train generative models based on Variational Auto-Encoders (VAEs) \cite{kingma_auto-encoding_2014} to map high-dimensional data in $\mathcal{X}$ to a low-dimensional latent space $\mathcal{Z}$ by approximating the posterior distribution $q(\mathbf{z} | \mathbf{x})$ (encoder). The latent vectors $\mathbf{z} \in \mathcal{Z}$ can then be sampled to generate new data in $\mathcal{X}$ using the learned likelihood $p(\mathbf{x} | \mathbf{z})$ (decoder). We use different supervised learning methods to enforce disentanglement in the latent space by regularizing specific attributes of interest along certain dimensions of the latent space. These attributes can then be manipulated by using simple traversals across the regularized dimensions. Once the models are trained, different experiments are conducted to evaluate disentanglement and controllability. 

  \subsection{Learning Methods}
    Three different disentanglement learning methods are considered. Each method adds a supervised regularization loss to the VAE-training objective 
    \begin{equation}
      \label{eq:loss_beta_vae}
      \mathnormal{L} = \mathnormal{L}_{\mathrm{VAE}} + \gamma \mathnormal{L}_\mathrm{reg},
    \end{equation}
    \noindent{where} $\mathnormal{L}$, $\mathnormal{L}_{\mathrm{VAE}}$, $\mathnormal{L}_\mathrm{reg}$ correspond to the overall loss, the VAE-loss \cite{kingma_auto-encoding_2014}, and the regularization loss respectively. The hyperparameter $\gamma$ is called the regularization strength.

    The first method, referred to as I-VAE, is based on the regularization proposed by Adel et al.~\cite{adel_discovering_2018}. It uses a separate linear classifier attached to each regularized dimension to predict the attribute classes. Note that while Adel et al.~use this regularization while learning a non-linear transformation of a latent space, we apply it during training of the latent space itself. This is a suitable choice for categorical attributes and is similar to the regularizer used in MIDI-VAE \cite{brunner_midi-vae_2018}. 
    The second method is the S2-VAE \cite{Locatello2020Disentangling}. This regularization, designed for continuous attributes, uses a binary cross-entropy loss to match attribute values to the regularized dimension. 
    The third method is the AR-VAE \cite{pati2020arvae}, which uses a batch-dependent regularization loss to encode continuous-valued attributes along specific dimensions of the latent space. This method is effective at regularizing note density and rhythm-based musical attributes \cite{tan2020music}. For comparison, baseline results obtained using the unsupervised $\beta$-VAE method \cite{higgins_beta-vae_2017} are also provided.
  
  \subsection{Dataset \& Data Representation}
    To conduct a systematic study and objectively evaluate the different methods, not only do we need to be able to measure the degree of disentanglement in the learned representations, but we should also be able to measure the attribute values in the generated data. Considering this, we use the dMelodies dataset \cite{pati2020dmelodies} which is an algorithmically constructed dataset with well-defined factors of variation specifically designed to enable objective evaluation of disentanglement learning methods for musical data. This dataset consists of simple 2-bar monophonic melodies which are based on arpeggiations over the standard I-IV-V-I cadence chord pattern. The dataset has the following factors of variation: \textit{Tonic}, \textit{Octave}, \textit{Scale}, \textit{Rhythm} for bars 1 and 2, and the \textit{Arpeggiation} directions for each of four chords. We use the tokenized data representation used by dMelodies \cite{pati2020dmelodies}. 
    
\begin{figure}[t]
  \centering
  \includegraphics[width = 0.8\columnwidth]{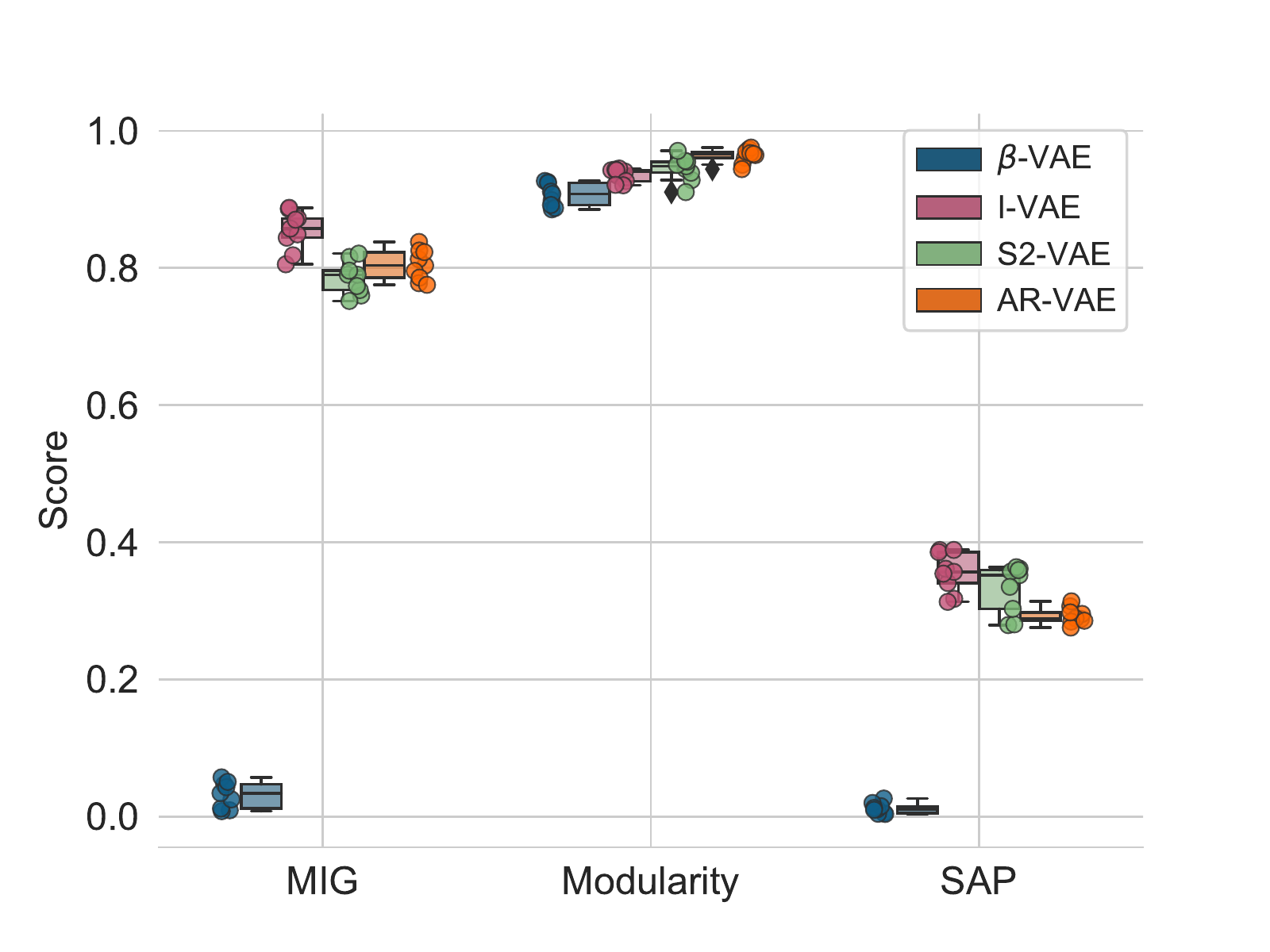}
  \caption[Overall disentanglement performance of different supervised methods on dMelodies]{Overall disentanglement performance (higher is better) of different supervised methods on dMelodies. Individual points denote results for different hyperparameter and random seed combinations.} 
\label{fig:sup_disent_results}
\end{figure}

\begin{figure*}[t]
  \centering
  \begin{tabular}{@{}c@{}}
      \includegraphics[width=0.23\textwidth]{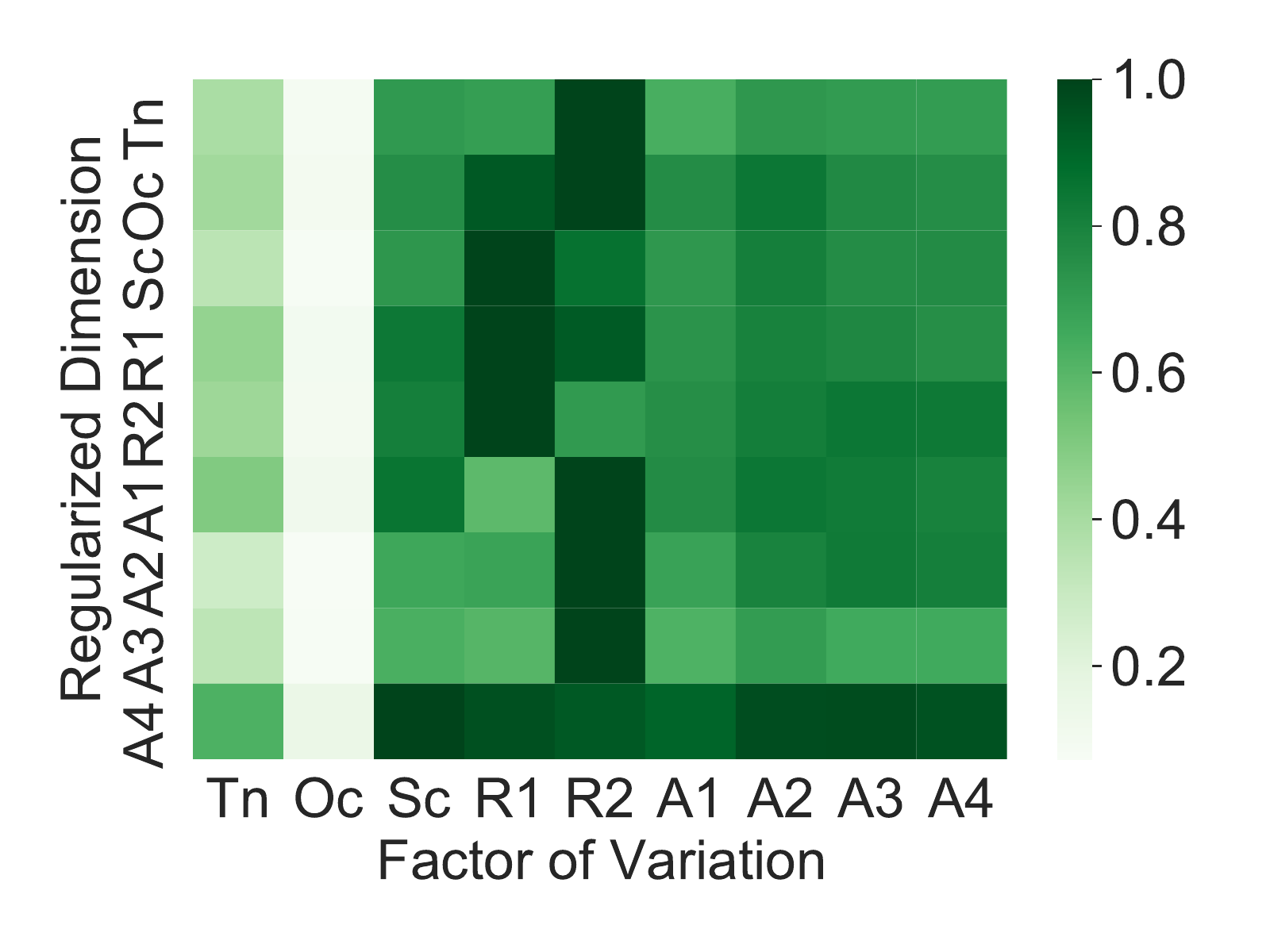}\vspace{-8pt} \\[\abovecaptionskip]
      \small (a) $\beta$-VAE
  \end{tabular}
  \begin{tabular}{@{}c@{}}
      \includegraphics[width=0.23\textwidth]{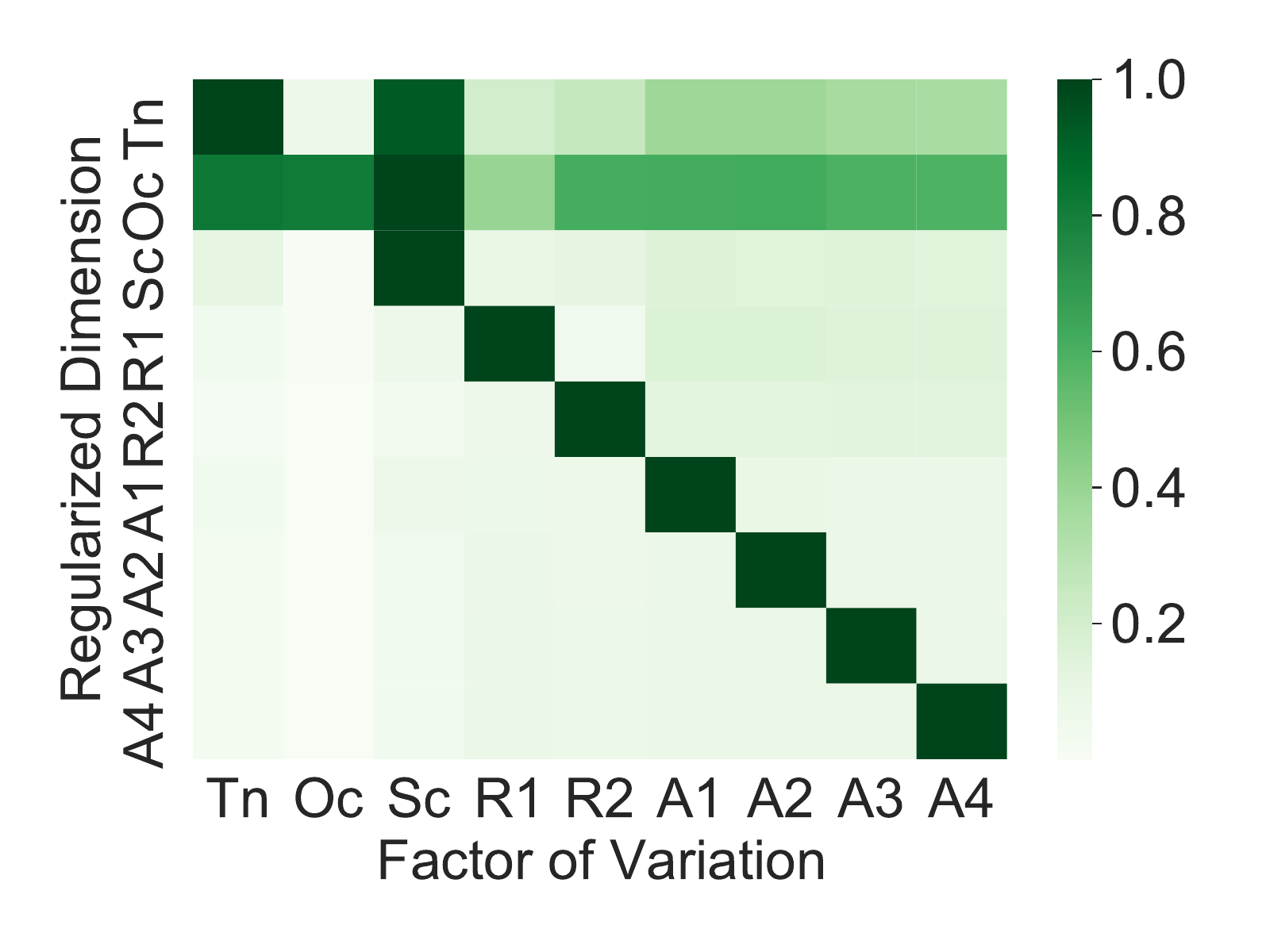}\vspace{-8pt} \\[\abovecaptionskip]
      \small (b) I-VAE
  \end{tabular}
  \begin{tabular}{@{}c@{}}
      \includegraphics[width=0.23\textwidth]{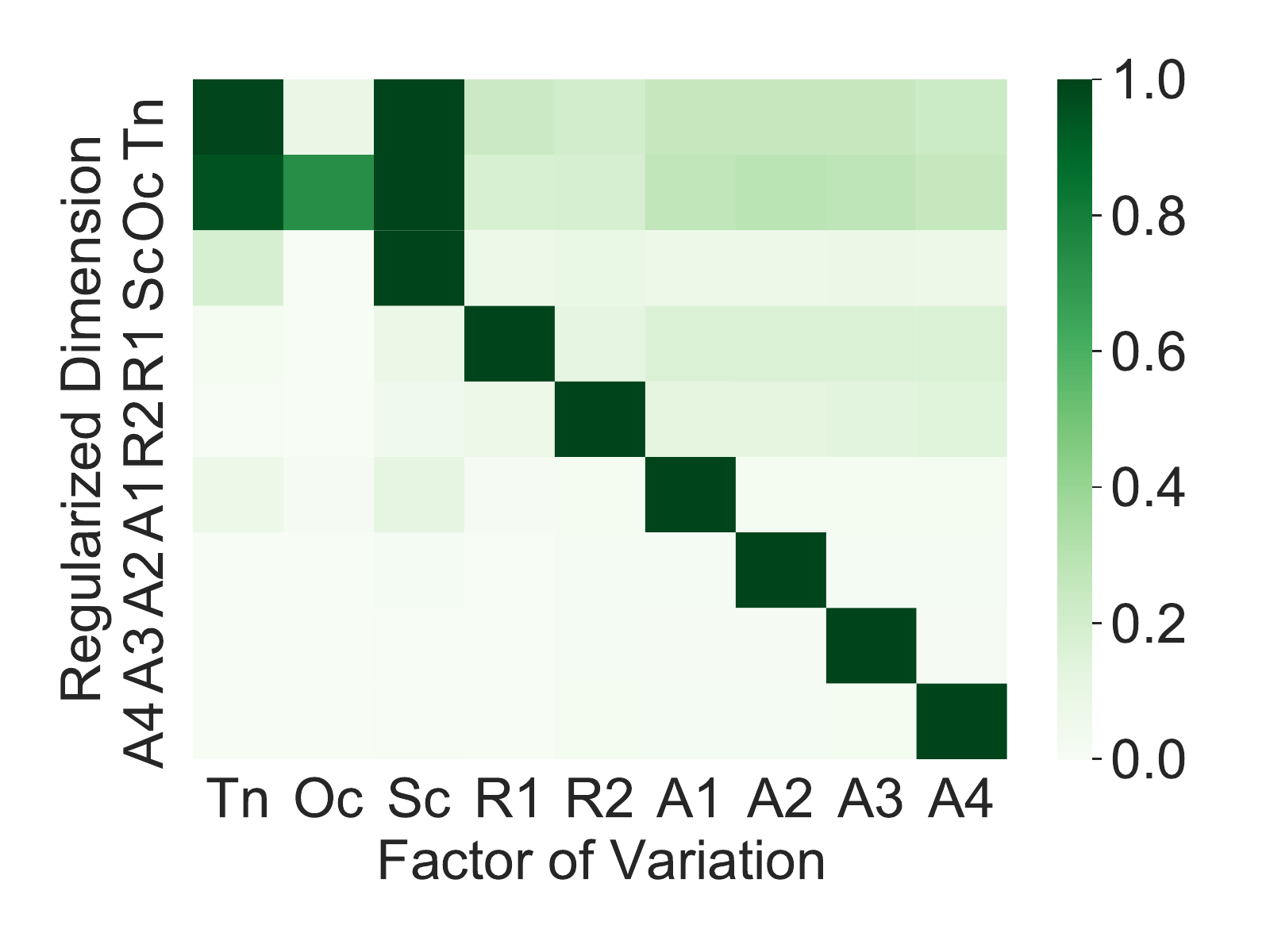}\vspace{-8pt} \\[\abovecaptionskip]
      \small (c) S2-VAE
  \end{tabular}
  \begin{tabular}{@{}c@{}}
      \includegraphics[width=0.23\textwidth]{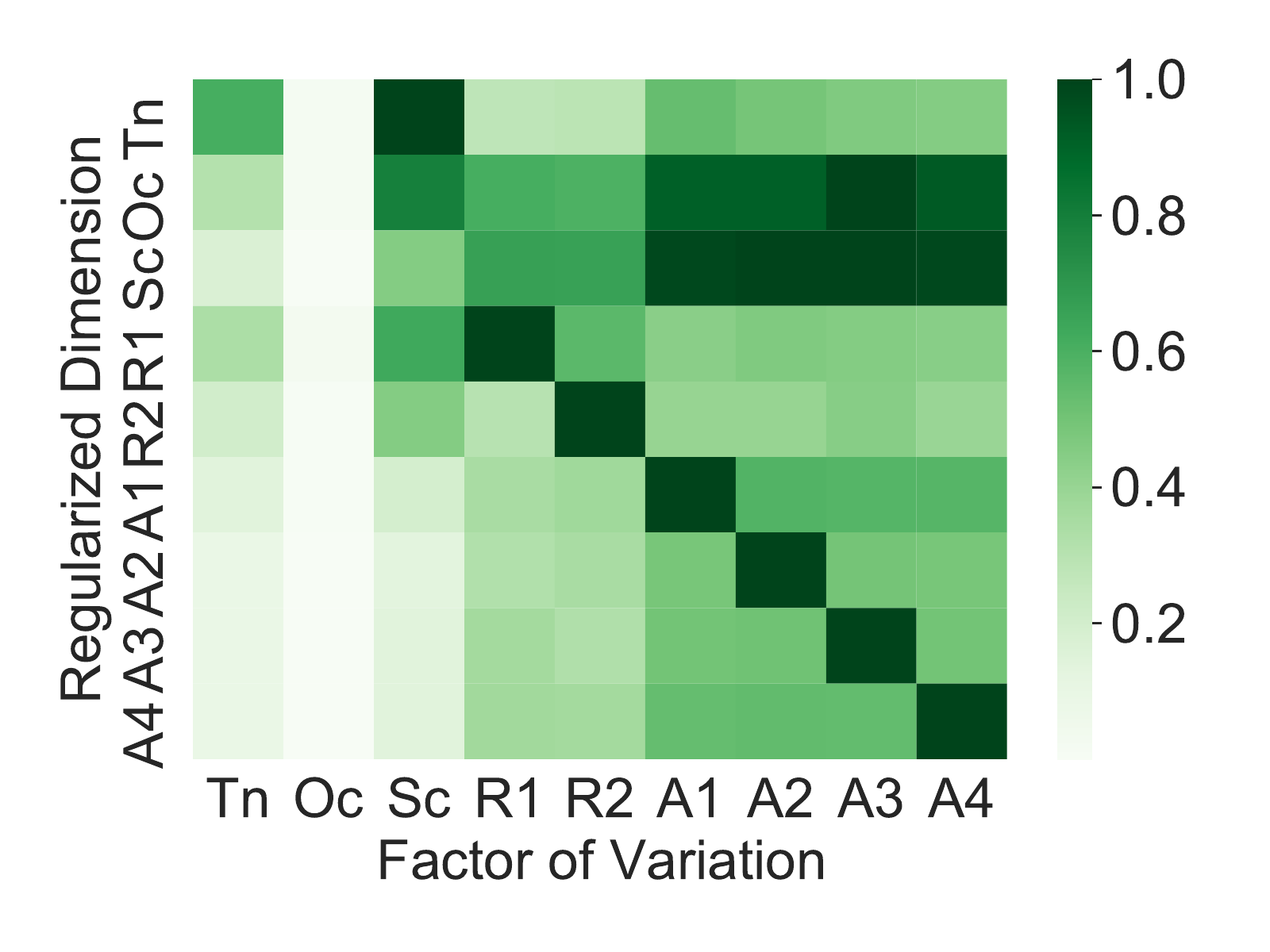}\vspace{-8pt} \\[\abovecaptionskip]
      \small (d) AR-VAE
  \end{tabular}
  \caption[Attribute-change matrices for different methods]{Attribute-change matrices for different methods. Tn: Tonic, Oc: Octave, Sc: Scale, R1 and R2: rhythm for bars 1 and 2 respectively, A1-A4: arpeggiation direction for the four chords.}
\label{fig:sup_attr_disent}
\end{figure*}

  \subsection{Model Architectures \& Training Specifications}
    The VAE architecture is based on a hierarchical RNN model \cite{pati2020dmelodies}, which is inspired by the MusicVAE model \cite{roberts_hierarchical_2018}. Additional experiments using a CNN-based architecture are omitted here for brevity but provided in the supplementary material. Since both S2-VAE and AR-VAE are designed for continuous attributes, the factors of variation are treated as continuous values by considering the index of the category as the attribute value and then normalizing them to $[0, 1]$. For instance, the \textit{Scale} attribute has $3$ distinct options and hence, the normalized continuous values are $[0, \frac{1}{2}, 1]$ corresponding to the major, harmonic minor, and blues scales, respectively. Three different values of regularization strength $\gamma \in \left\{ 0.1, 1.0, 10.0 \right\}$ are used.

    For each of the above methods and hyperparameter combinations, three models with different random seeds are trained. The dataset is divided into training, validation, and test set using a $70\%$-$20\%$-$10\%$ split. To ensure consistency across training, all models are trained with a batch size of $512$ for $100$ epochs. The ADAM optimizer \cite{kingma_adam_2015} is used with a fixed learning rate of $1\mbox{e\ensuremath-}4$, $\beta _{1}=0.9$, $\beta _{2}=0.999$, and $\epsilon = 1\mbox{e\ensuremath-}8$.

\section{Results and Discussion} \label{sec:results}
  We now present and discuss the results of the different experiments conducted. The first experiment objectively measures the degree of disentanglement in the representations learned using the different methods. The second experiment evaluates the degree to which each method allows independent control over the different attributes. The third experiment throws additional light into the behavior by visualizing the latent spaces with respect to the different attributes. Then, we introduce a new metric to evaluate the quality of latent spaces with respect to the decoder. Finally, we present a qualitative inspection of the data generated by traversals along different regularized dimensions to further illustrate the key findings.  
  
  \subsection{Attribute Disentanglement} \label{dis:sup_disent}
  
    In order to objectively measure disentanglement, we rely on commonly used metrics: 
    \begin{inparaenum}[(a)]
        \item Mutual Information Gap (MIG) \cite{chen_isolating_2018}, which measures the difference of mutual information between a given attribute and the top two dimensions of the latent space that share maximum mutual information with the attribute, 
        \item Modularity \cite{ridgeway_learning_2018}, which measures if each dimension of the latent space depends on only one attribute, and 
        \item Separated Attribute Predictability (SAP) \cite{kumar_variational_2017}, which measures the difference in the prediction error of the two most predictive dimensions of the latent space for a given attribute. 
    \end{inparaenum}
    For each metric, the mean across all attributes is used for aggregation. For consistency, standard implementations are used \cite{locatello_challenging_2019}.  

    The disentanglement performance of the three supervised methods on the held-out test set is compared against the $\beta$-VAE model in \figref{fig:sup_disent_results}. Unsurprisingly, all three supervised methods outperform the $\beta$-VAE across the three disentanglement metrics. The improvement is much higher for the \textit{MIG} and \textit{SAP} score which both measure the degree to which each attribute is encoded only along a single dimension of the latent space. %

    Using supervision, therefore, leads to better overall disentanglement. Note that this superior performance is achieved without sacrificing the reconstruction quality. All three supervised methods achieve a reconstruction accuracy $> 90\%$. This is a considerable improvement over the unsupervised learning methods seen in the dMelodies benchmarking experiments (average accuracy of $\approx 50\%$ \cite{pati2020arvae}).

  \subsection{Independent Control during Generation} \label{dis:sup_attr_disent}
    Considering that supervised methods can obtain better disentanglement along with good reconstruction accuracy, we now look at how effective these methods are for independently controlling different attributes. To measure this quantitatively, we propose the following protocol. Given a data-point with latent vector $\mathbf{z}$, $6$ different variations are generated by uniformly interpolating along the dimension $r_{l}$, where $r_{l}$ is the regularized dimension for attribute $a_l$. The limits of interpolation are chosen based on the maximum and minimum latent code values obtained during encoding the validation data. For the $\beta$-VAE model, the dimension with the highest mutual information with the attribute is considered as the regularized dimension. An attribute change matrix $A \in \mathbb{R}^{L \times L}$, where $L$ is the number of attributes, is computed using the following formulation:
    \begin{equation}
      A(m,n)  = \sum_{i=1}^{6} \big [0 \neq   \left |  a_{n}(\mathbf{z}_{i}^{m}) - a_{n}(\mathbf{z})   \right | \big ]     ,    
    \end{equation}
    \noindent{where} $A(m,n)$ computes the net change in the $n^{\mathrm{th}}$ attribute as one traverses the dimension $r_m$ (which regularizes the $m^{\mathrm{th}}$ attribute), $\left[ \cdot \right]$ represents the inverse Kronecker delta function, $a_{n}(\cdot)$ is the value of the $n^{\mathrm{th}}$ attribute, and $\mathbf{z}_{i}^{m}$ is the $i^{\mathrm{th}}$ interpolation of $\mathbf{z}$ obtained by traversing along the $r_{m}$ dimension. This attribute change matrix is computed for each model type by averaging over a total of $1024$ data-points in the test-set and across all $3$ random seeds (regularization hyperparameters are fixed at $\beta=0.2$, $\gamma=1.0$). The matrix is also normalized so that the maximum value across each row corresponds to one. Independent control over attributes should result in the matrix $A$ having high values along the diagonal and low values on the off-diagonal entries which would denote that traversing a regularized dimension only affects the regularized attribute. 
                
    The following observations can be made from the matrices visualized in \figref{fig:sup_attr_disent}. First, $\beta$-VAE performs the worst as traversals along different dimensions change multiple attributes simultaneously. Second, among the supervised methods, I-VAE and S2-VAE seem to perform better than AR-VAE. This can be seen from the lighter shades of the off-diagonal elements in the plots for I-VAE and S2-VAE. While the better performance of I-VAE is expected since it is designed for categorical attributes, the poorer performance of AR-VAE in comparison to S2-VAE needs further investigation. Finally, the \textit{scale} attribute (3rd column) changes the most while traversing the regularized dimensions for the supervised methods. This indicates that all supervised methods struggle in generating notes conforming to particular scales. One explanation for this could be that the \textit{scale} is the most complex among all the attributes. Note that while there is no considerable difference between the disentanglement performance of the three methods (compare \figref{fig:sup_disent_results}), I-VAE and S2-VAE show much better performance compared to AR-VAE in this experiment which shows that disentanglement does not ensure better controllability.

    \begin{figure}[t]
      \centering
      \begin{tabular}{@{}c@{}}
        \includegraphics[width=0.23\textwidth]{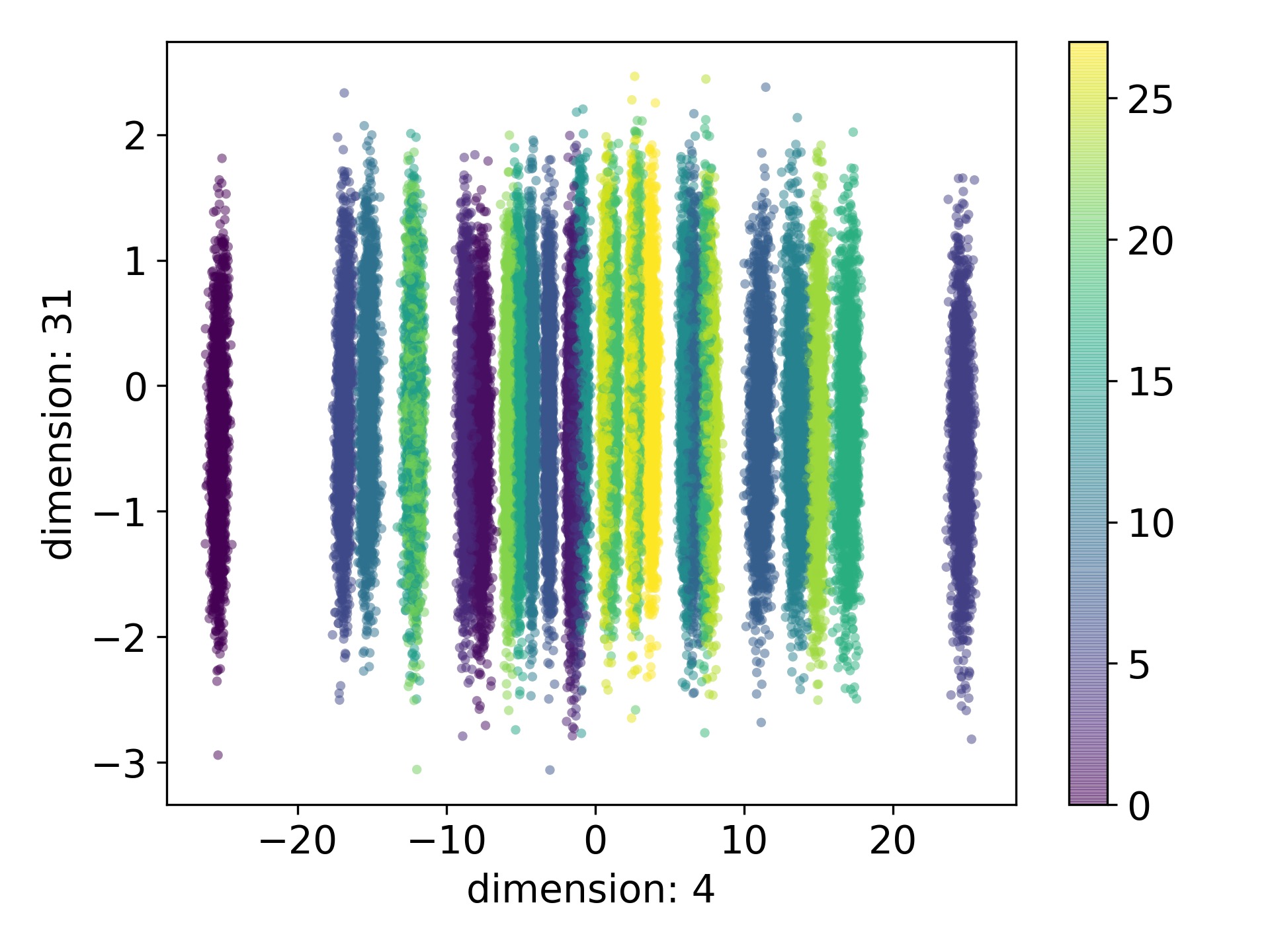}\vspace{-8pt} \\[\abovecaptionskip]
      \end{tabular}
      \begin{tabular}{@{}c@{}}
        \includegraphics[width=0.23\textwidth]{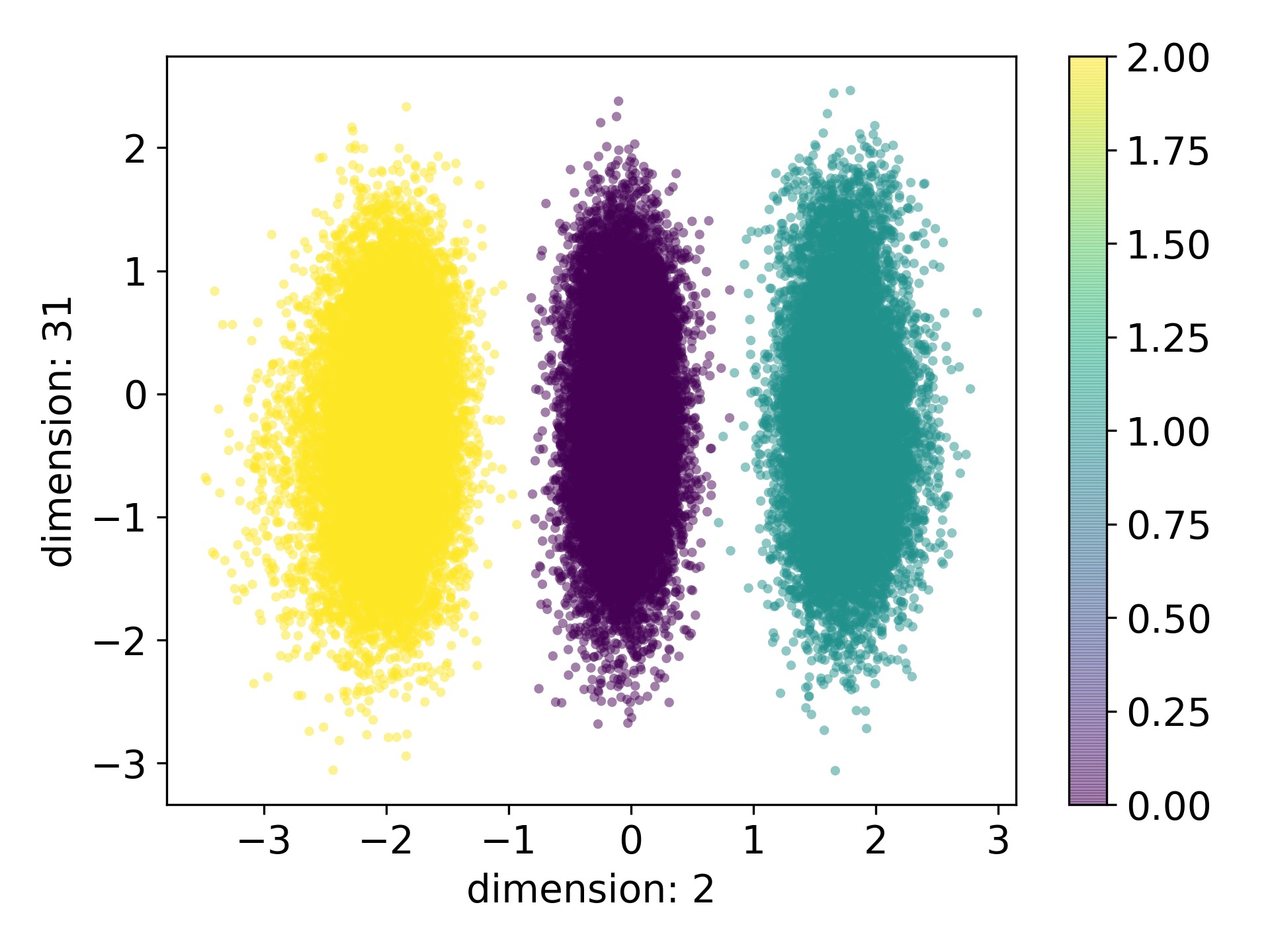}\vspace{-8pt} \\[\abovecaptionskip]
      \end{tabular}
      
      \begin{tabular}{@{}c@{}}
        \includegraphics[width=0.23\textwidth]{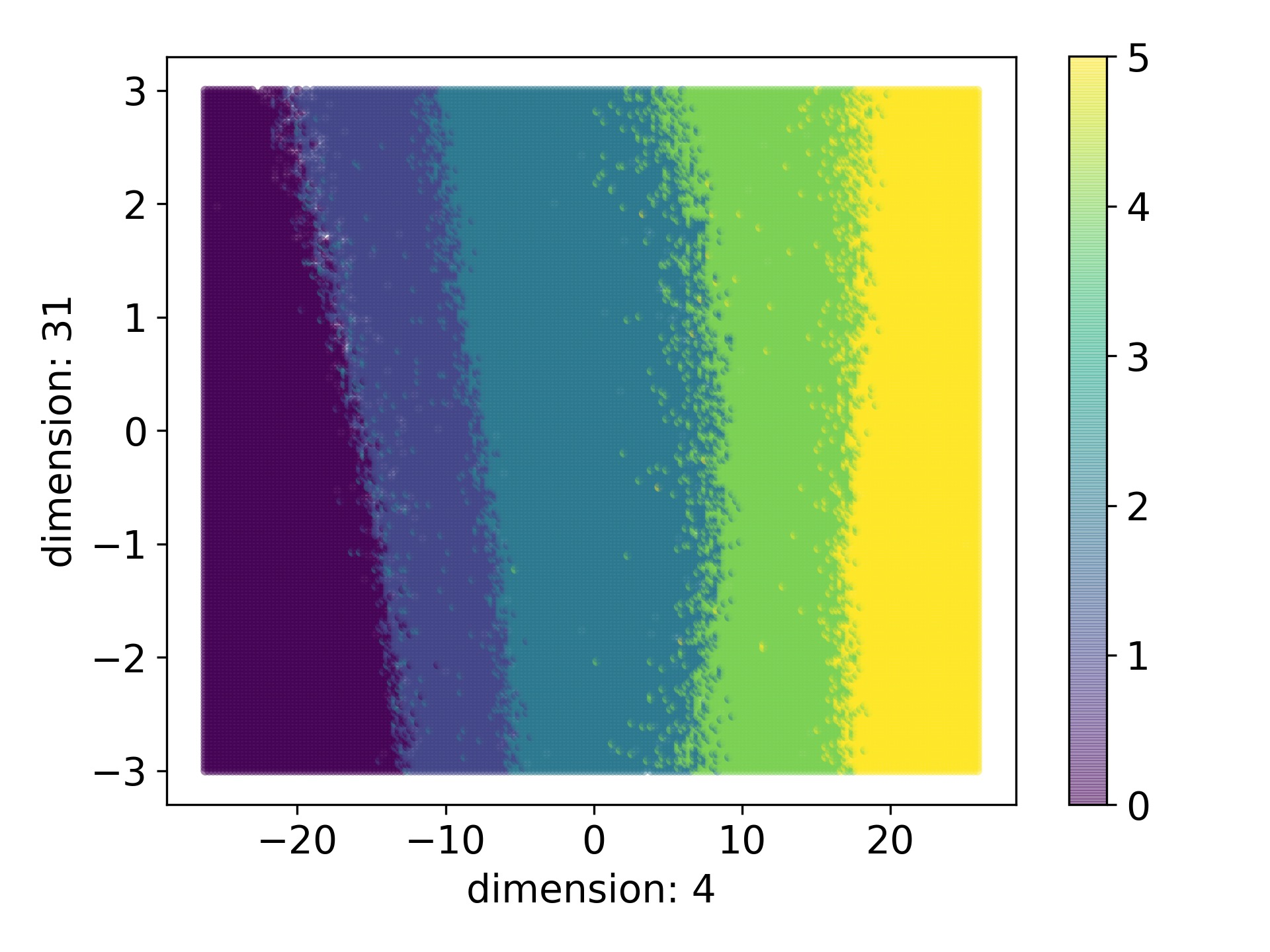}\vspace{-8pt} \\[\abovecaptionskip]
        \small (a) \textit{Rhythm Bar 2}
      \end{tabular}
      \begin{tabular}{@{}c@{}}
        \includegraphics[width=0.23\textwidth]{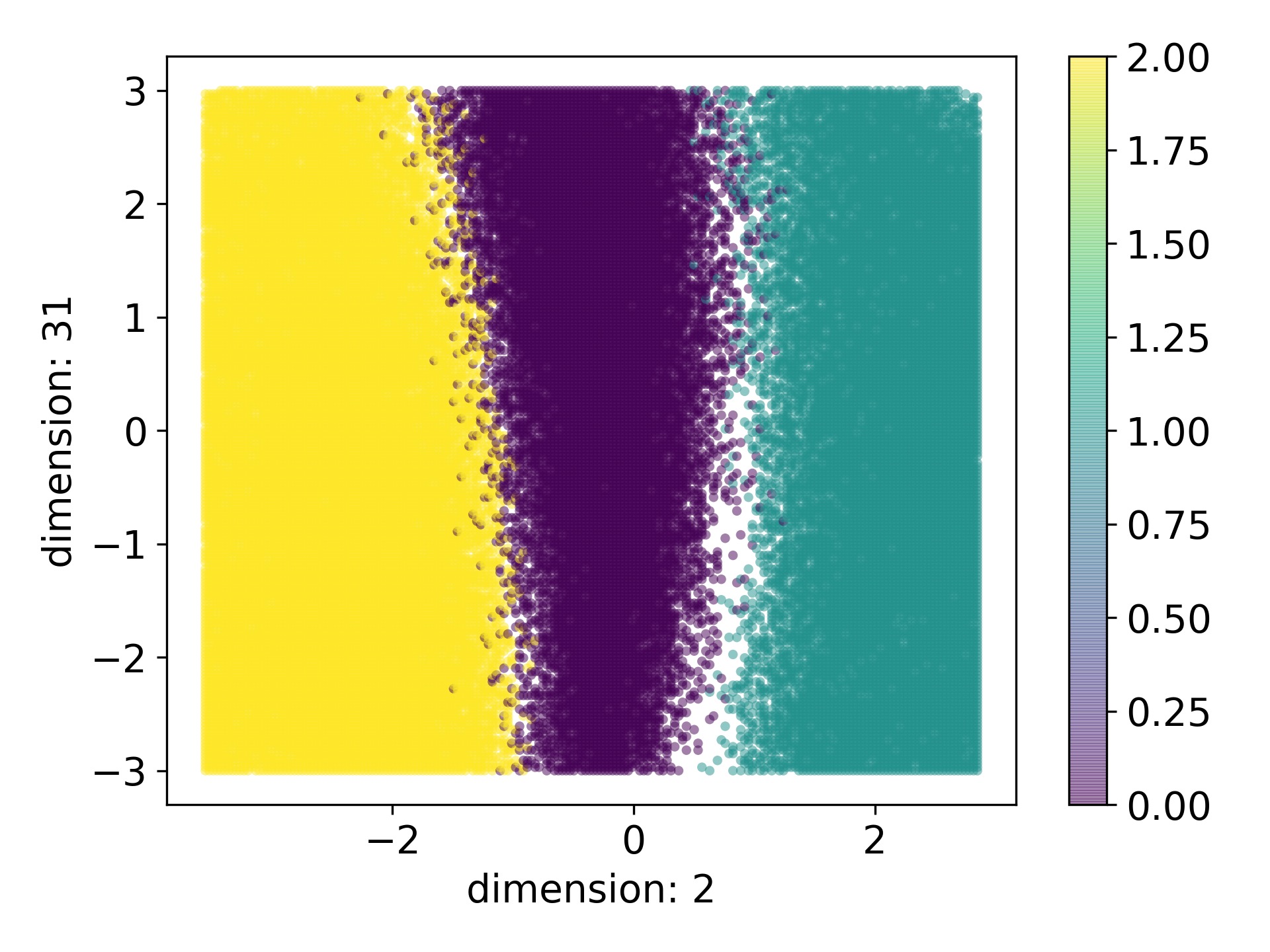}\vspace{-8pt} \\[\abovecaptionskip]
        \small (b) \textit{Scale}
      \end{tabular}

      \caption[Data distribution and surface plots for I-VAE]{Data distribution (top row) and surface plots (bottom row) for I-VAE.} 
    \label{fig:dis_viz_IVAE}
    \end{figure}
  
  \subsection{Latent Space Visualization} \label{dis:sup_latent_viz}

    To better understand the difference between disentanglement and controllability of attributes, we try to visualize the structure of the latent space with respect to the different attributes. This is done using 2-dimensional \textit{data distribution} and \textit{latent surface} plots. Both plots show the variance of a given attribute (using different colors for different attribute values) with respect to the regularized dimension (shown on the $x$-axis) and a randomly chosen non-regularized dimension (shown on the $y$-axis). 
    
    For the data distribution plots, first, latent representations are obtained for data in the held-out test set using the VAE-encoder. Then, for each attribute, these representations are projected onto a 2-dimensional plane where the \textit{x}-axis corresponds to the regularized dimension and the \textit{y}-axis corresponds to a non-regularized dimension. To generate the surface plots, for a given attribute, a 2-dimensional plane on the latent space is considered which comprises the regularized dimension for the attribute and a non-regularized dimension. The latent code for the other dimensions is drawn from a normal distribution and kept fixed. The latent vectors thus obtained are passed through the VAE decoder and the attributes of the generated data are plotted.

    \begin{figure}[t]
      \centering
      \begin{tabular}{@{}c@{}}
        \includegraphics[width=0.23\textwidth]{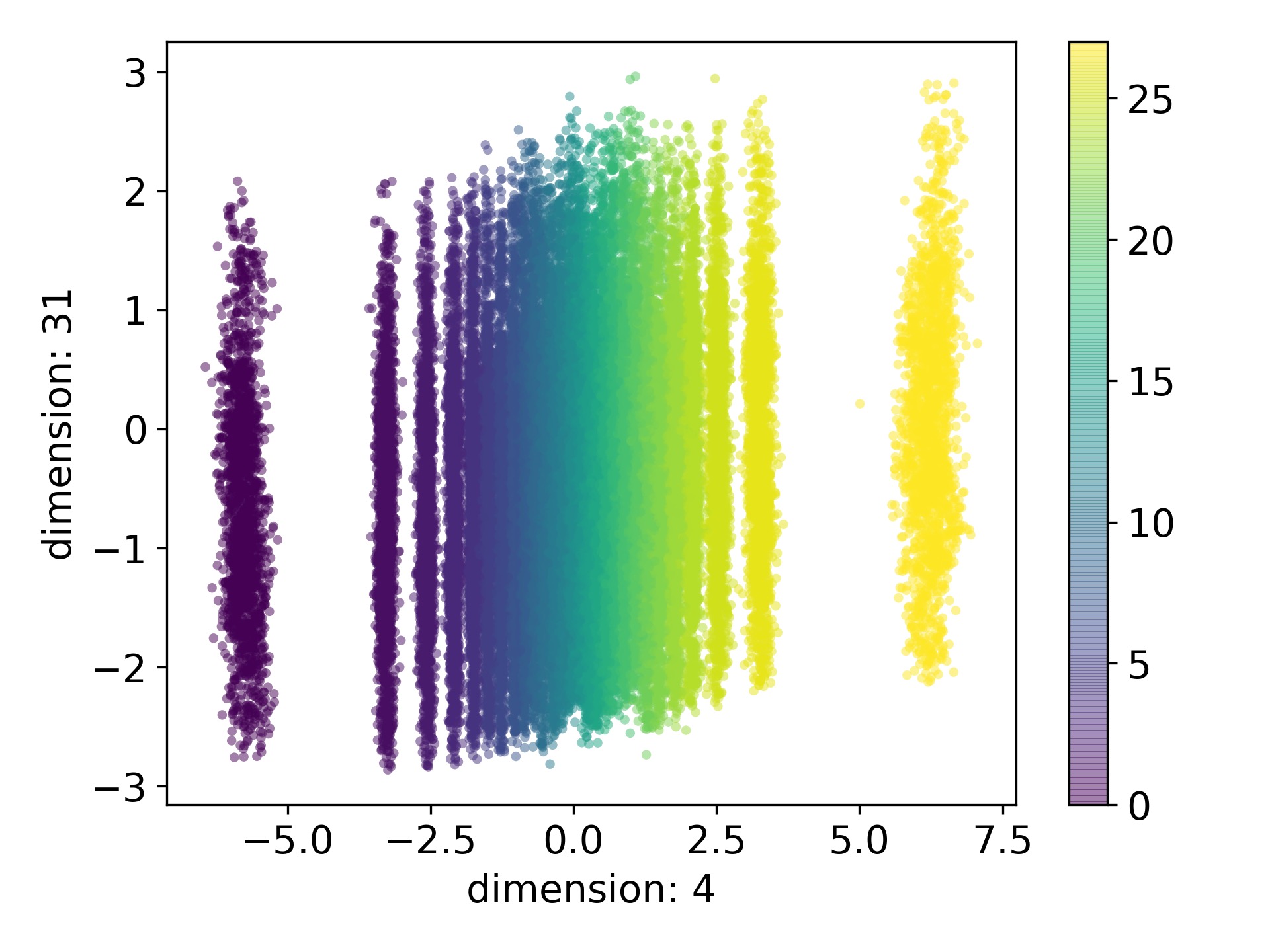}\vspace{-8pt} \\[\abovecaptionskip]
      \end{tabular}
      \begin{tabular}{@{}c@{}}
        \includegraphics[width=0.23\textwidth]{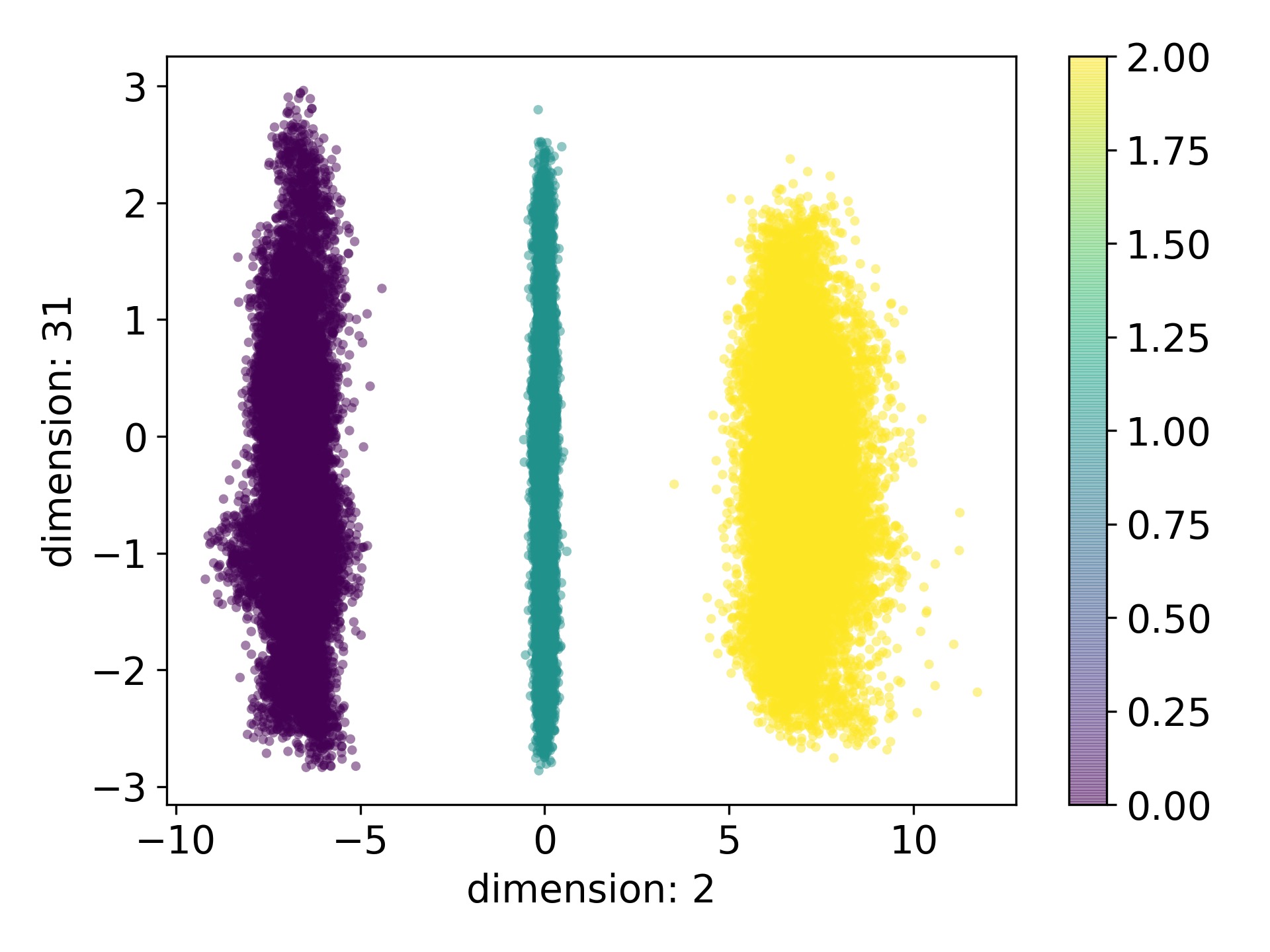}\vspace{-8pt} \\[\abovecaptionskip]
      \end{tabular}
      
      \begin{tabular}{@{}c@{}}
        \includegraphics[width=0.23\textwidth]{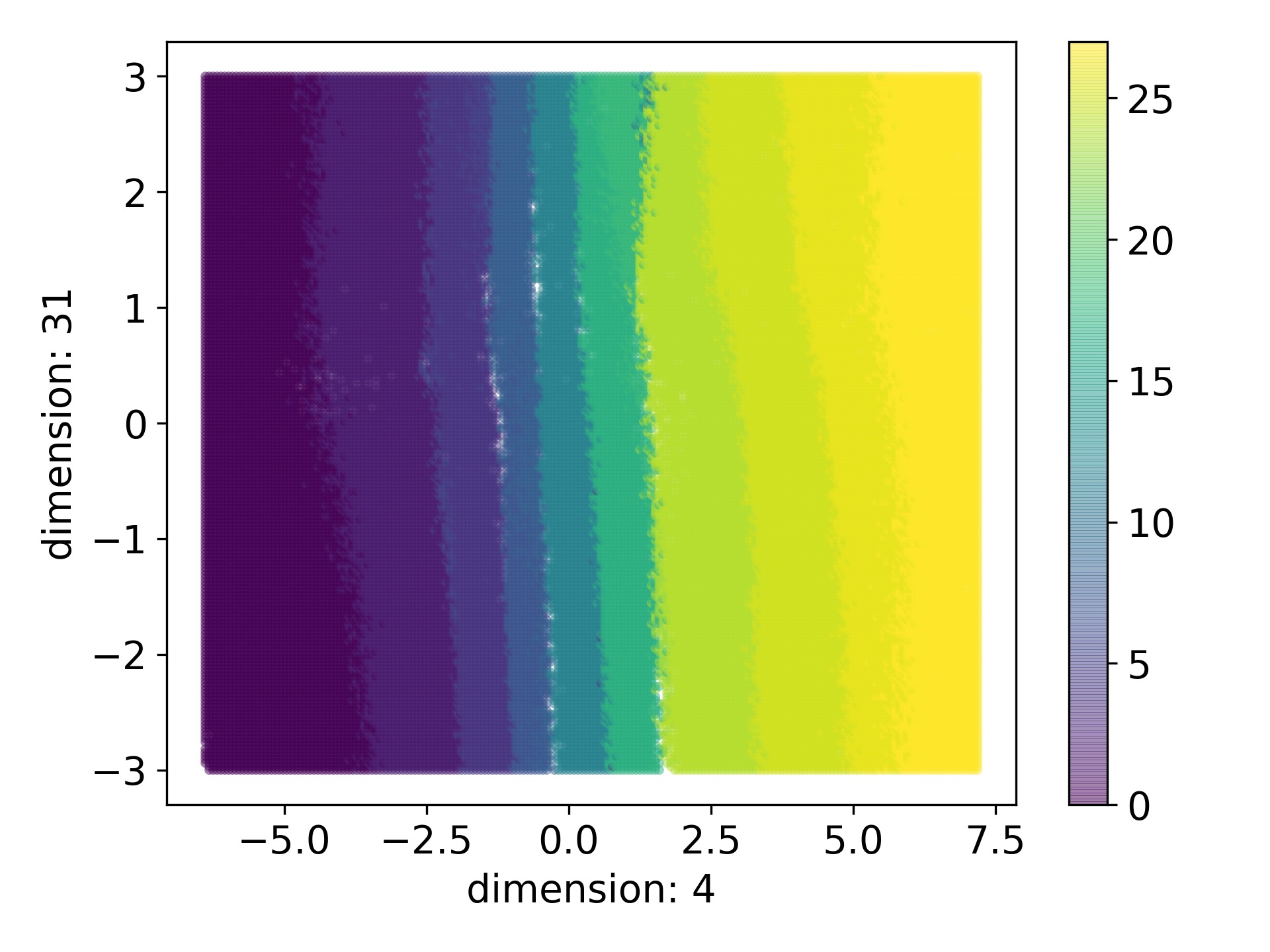}\vspace{-8pt} \\[\abovecaptionskip]
        \small (a) \textit{Rhythm Bar 2}
      \end{tabular}
      \begin{tabular}{@{}c@{}}
        \includegraphics[width=0.23\textwidth]{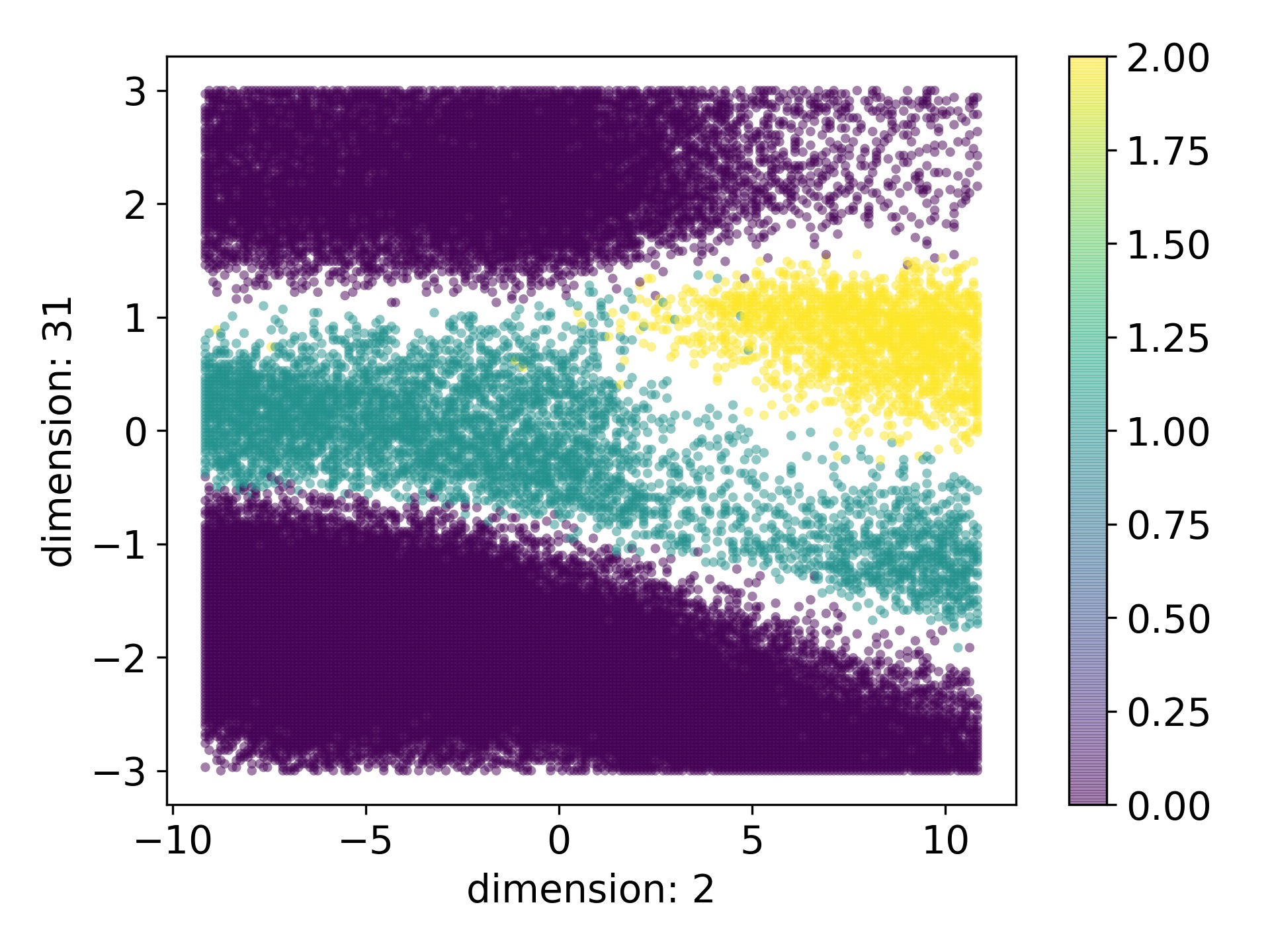}\vspace{-8pt} \\[\abovecaptionskip]
        \small (b) \textit{Scale}
      \end{tabular}
      
      \caption[Data distribution and surface plots for the S2-VAE]{Data distribution (top row) and surface plots (bottom row) for S2-VAE.} 
    \label{fig:dis_viz_S2VAE}
    \end{figure}

\begin{figure}[t]
      \centering
      \begin{tabular}{@{}c@{}}
        \includegraphics[width=0.23\textwidth]{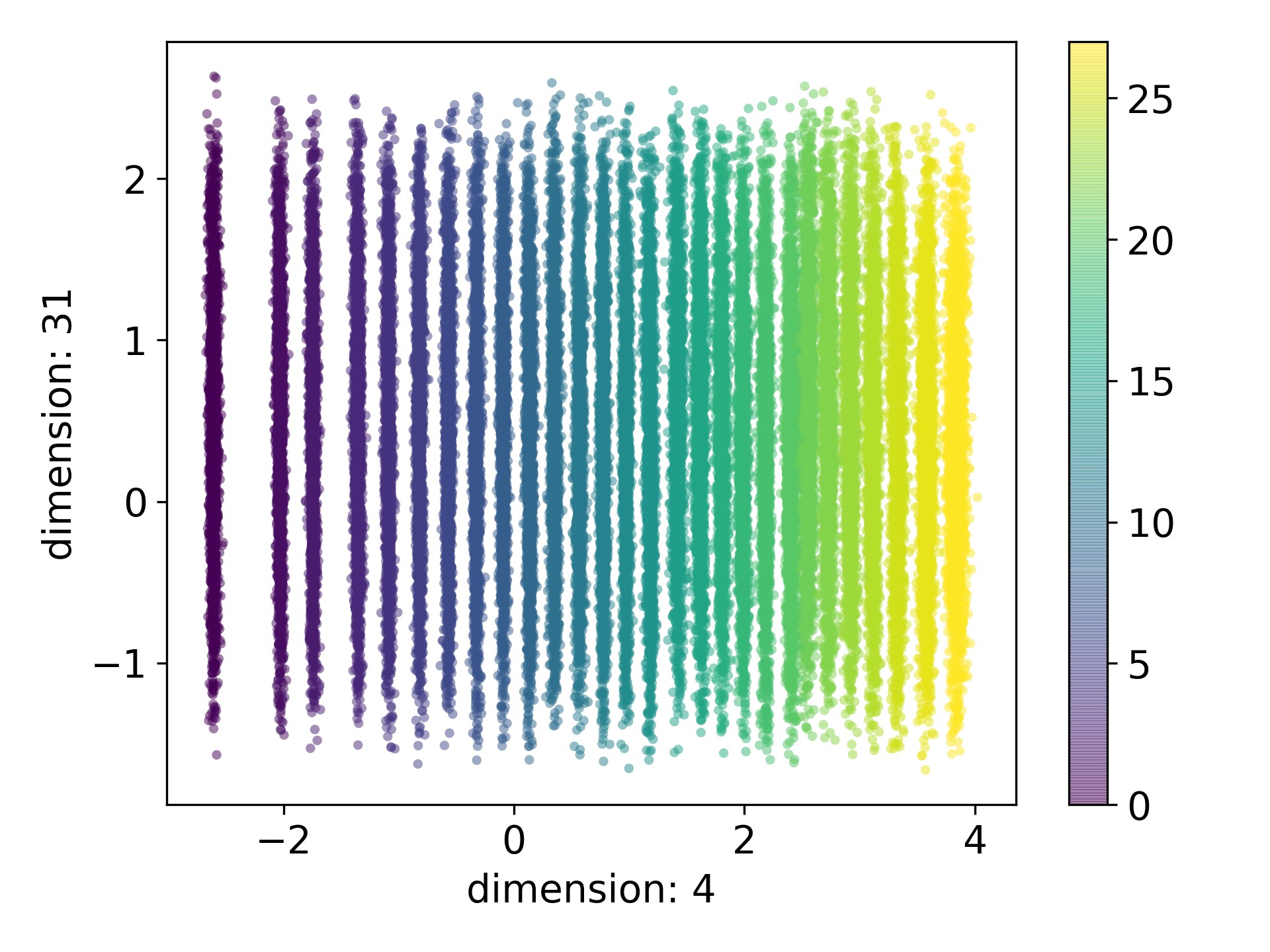}\vspace{-8pt} \\[\abovecaptionskip]
      \end{tabular}
      \begin{tabular}{@{}c@{}}
        \includegraphics[width=0.23\textwidth]{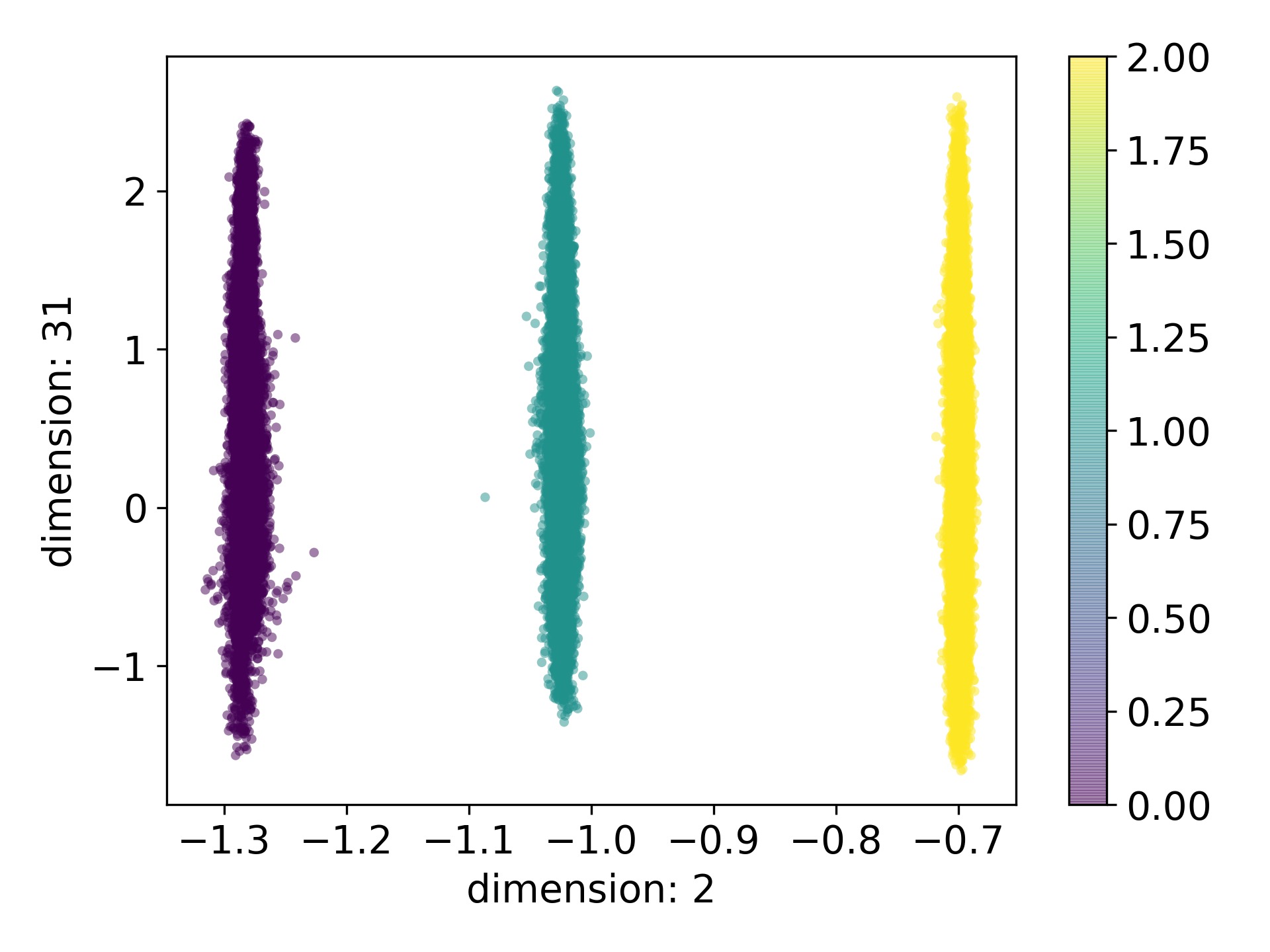}\vspace{-8pt} \\[\abovecaptionskip]
      \end{tabular}
      
      \begin{tabular}{@{}c@{}}
        \includegraphics[width=0.23\textwidth]{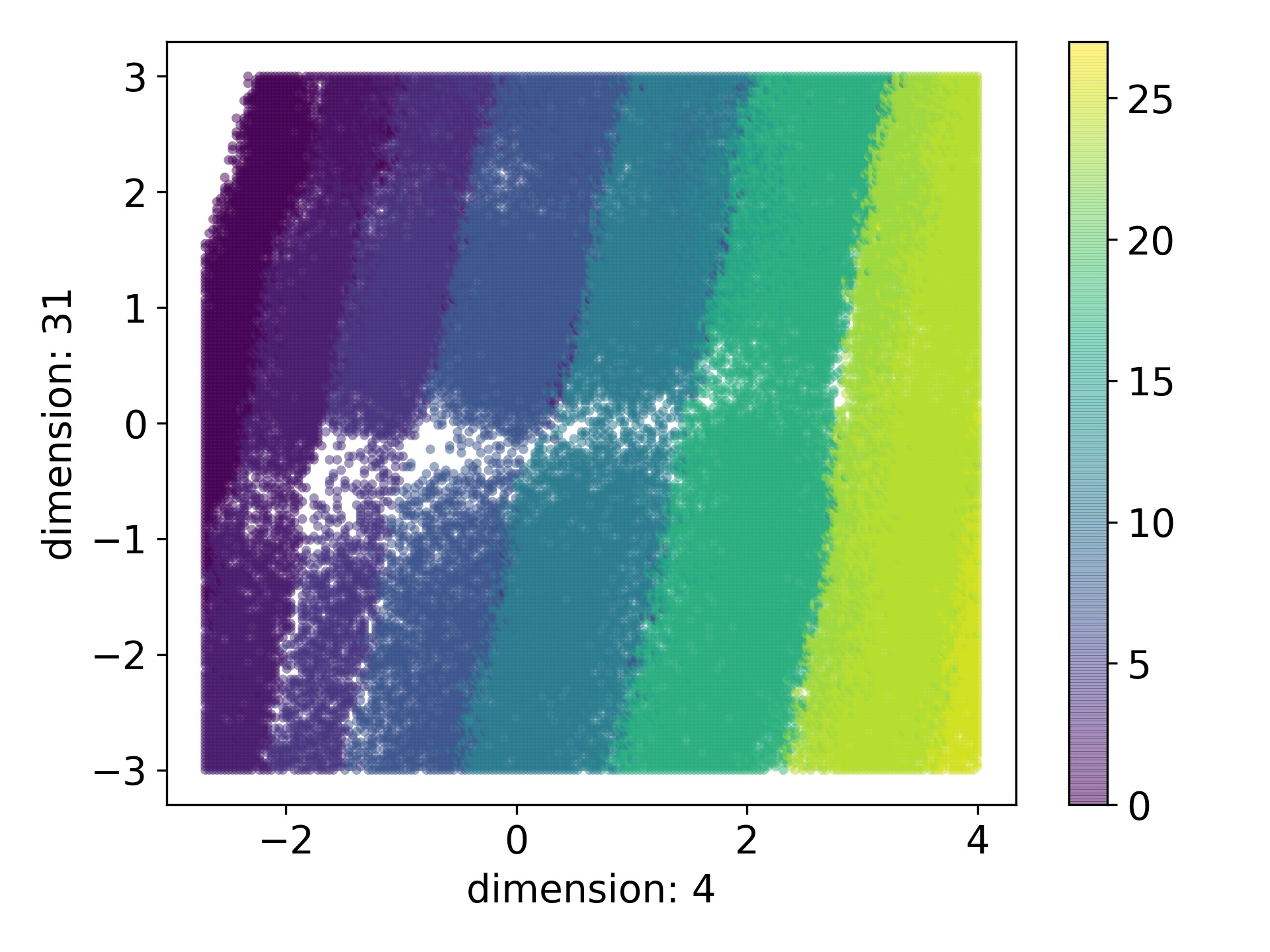}\vspace{-8pt} \\[\abovecaptionskip]
        \small (a) \textit{Rhythm Bar 2}
      \end{tabular}
      \begin{tabular}{@{}c@{}}
        \includegraphics[width=0.23\textwidth]{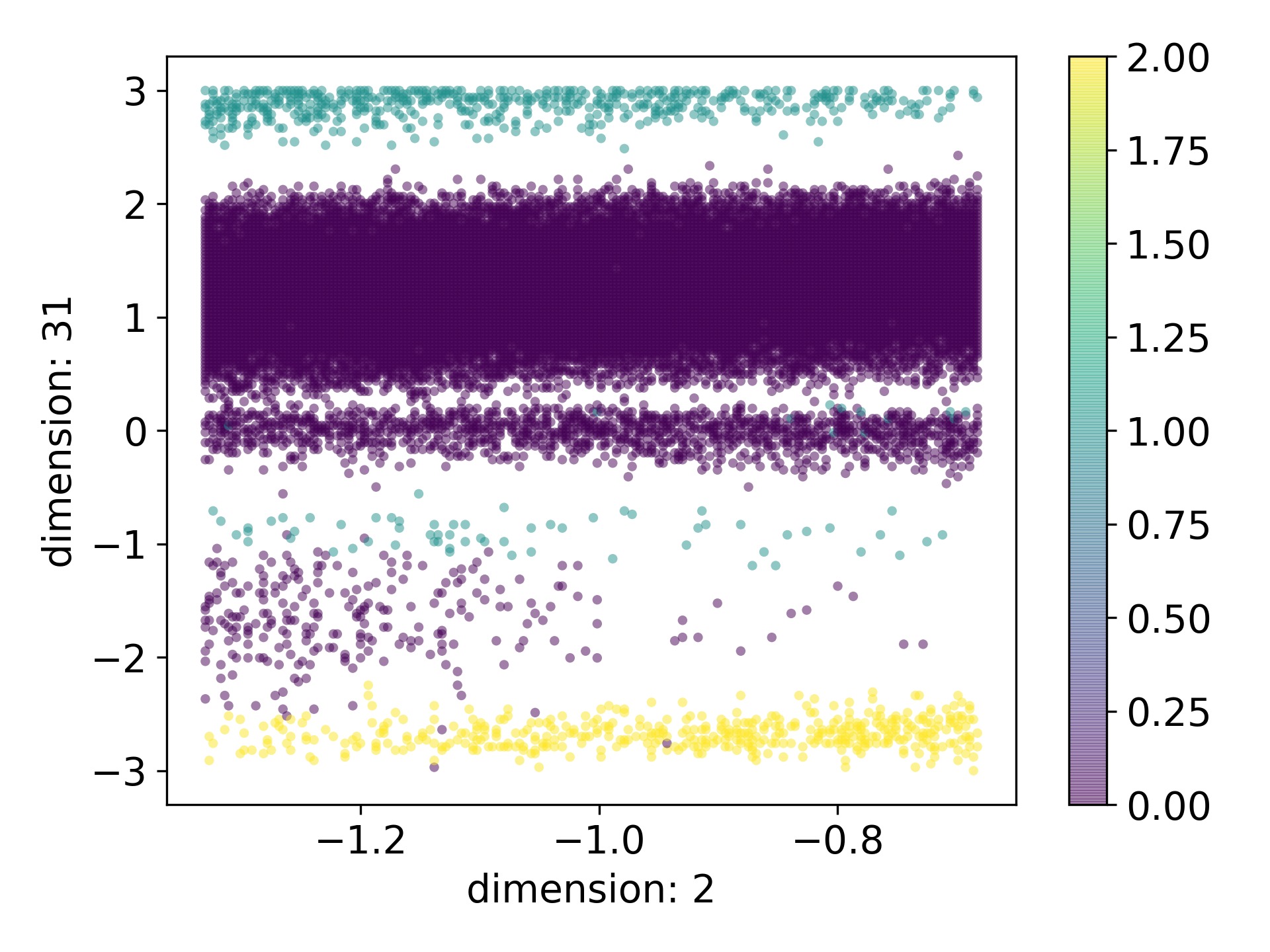}\vspace{-8pt} \\[\abovecaptionskip]
        \small (b) \textit{Scale}
      \end{tabular}

      \caption[Data distribution and surface plots for AR-VAE method]{Data distribution (top row) and surface plots (bottom row) for AR-VAE.} 
    \label{fig:dis_viz_ARVAE}
    \end{figure}

    Figures~\ref{fig:dis_viz_IVAE}, \ref{fig:dis_viz_S2VAE}, and \ref{fig:dis_viz_ARVAE} show the results for I-VAE, S2-VAE, and AR-VAE respectively. In each figure, the top row corresponds to the data distribution plots, and the bottom row shows the latent surface plots. For the surface plots, the generated data-points sometimes have attribute values that are either not present in the training set or cannot be determined (e.g., the generated melody might not conform to any of the $3$ possible scales in the dataset, or the arpeggiation direction might be neither up nor down). These \textit{undefined} or out-of-distribution attribute values are shown as empty spaces in the latent surface plots. 
           
    For all three methods, the data distribution plots (top rows) show a clear separation of attribute values along the regularized dimension which explains the high disentanglement performance seen in \secref{dis:sup_disent}. However, the methods differ considerably when the latent surface plots (bottom rows) are compared. 
    I-VAE (see \figref{fig:dis_viz_IVAE}) shows good performance where moving along the regularized dimension (\textit{x}-axis) changes the corresponding attribute, while traversals along the non-regularized dimension (\textit{y}-axis) have little effect. However, the manner of change is unpredictable. For instance, in \figref{fig:dis_viz_IVAE}(a)(bottom), only 5 out of the 28 possible rhythms are generated. In addition, the order of the generated rhythms is different from the encoder distribution in \figref{fig:dis_viz_IVAE}(a)(top). 
    In contrast, for S2-VAE, the gradual change of color in \figref{fig:dis_viz_S2VAE}(a)(bottom) shows a high degree of controllability for the rhythm attribute. However, it struggles to control the \textit{scale} attribute. Traversing along the non-regularized dimension in \figref{fig:dis_viz_S2VAE}(b)(bottom) results in an undesirable change in the \textit{scale} of the generated melody. The latent space of AR-VAE (see \figref{fig:dis_viz_ARVAE}) has the most discrepancies. Not only is the latent space not centered around the origin (see the top row of \figref{fig:dis_viz_ARVAE}(b)) for the  \textit{scale} attribute, but the degree of controllability is also poor. For instance, the \textit{scale} attribute does not change at all along the regularized dimension (see \figref{fig:dis_viz_ARVAE}(b)(bottom)). In addition, the empty spaces in the surface plots show that many of the generated data-points have an out-of-distribution attribute value. Results for all other attributes are provided in the supplementary material.

    The empty regions in the latent spaces show that while these methods can train strong discriminative encoders which are good for disentanglement, they tend to have weak generative decoders which are incapable of utilizing the learned disentangled representations thereby resulting in \textit{holes} or vacant regions in the latent space where the behavior of the decoder is unpredictable.

  \subsection{Latent Density Ratio} \label{dis:ldm}
    From the perspective of the VAE-decoder, holes in the latent space can have a significant impact on controllability. Yet, established metrics do not capture this phenomenon properly. To help quantify this, we propose the Latent Density Ratio (LDR) metric. We first sample a set of $N$ (=$10$k) points in the latent space, pass them through the VAE decoder, and compute the percentage of data-points with valid attribute values out of the total number $N$.
    The overall LDR is obtained by averaging this metric across all attributes. The results in \tabref{tab:ldm} show that both S2-VAE and I-VAE have a lower degree of holes (higher LDR value) in comparison to AR-VAE which is in line with observations in the previous experiments.
  
  \subsection{Qualitative Inspection of Latent Interpolations} \label{dis:sup_latent_interp}
\begin{figure*}[t]
      \centering
      \begin{tabular}{@{}c@{}}
        \includegraphics[width=0.23\textwidth]{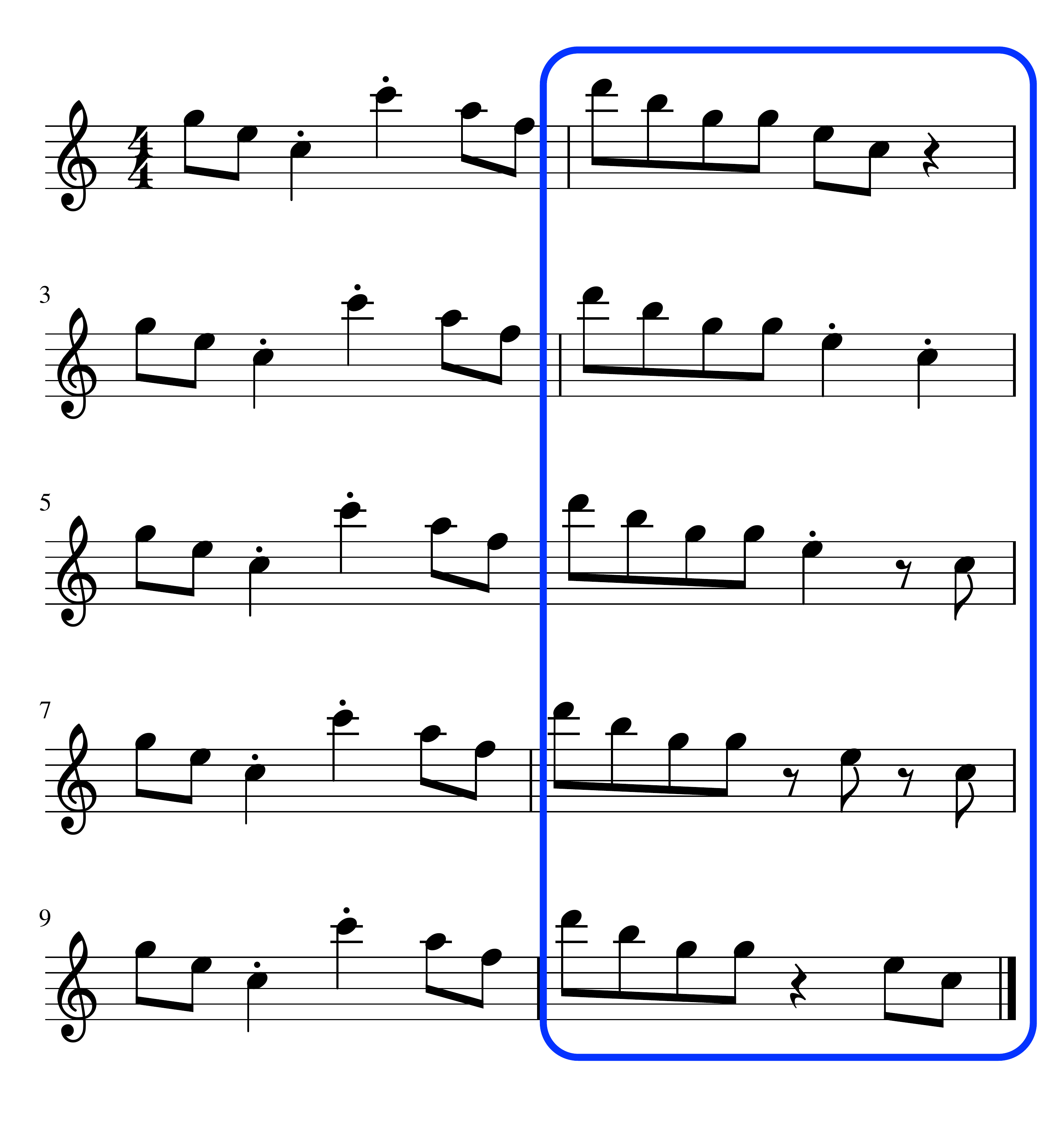}\vspace{-8pt} \\[\abovecaptionskip]
        \small (a) Rhythm Bar 2
      \end{tabular}
      \begin{tabular}{@{}c@{}}
        \includegraphics[width=0.23\textwidth]{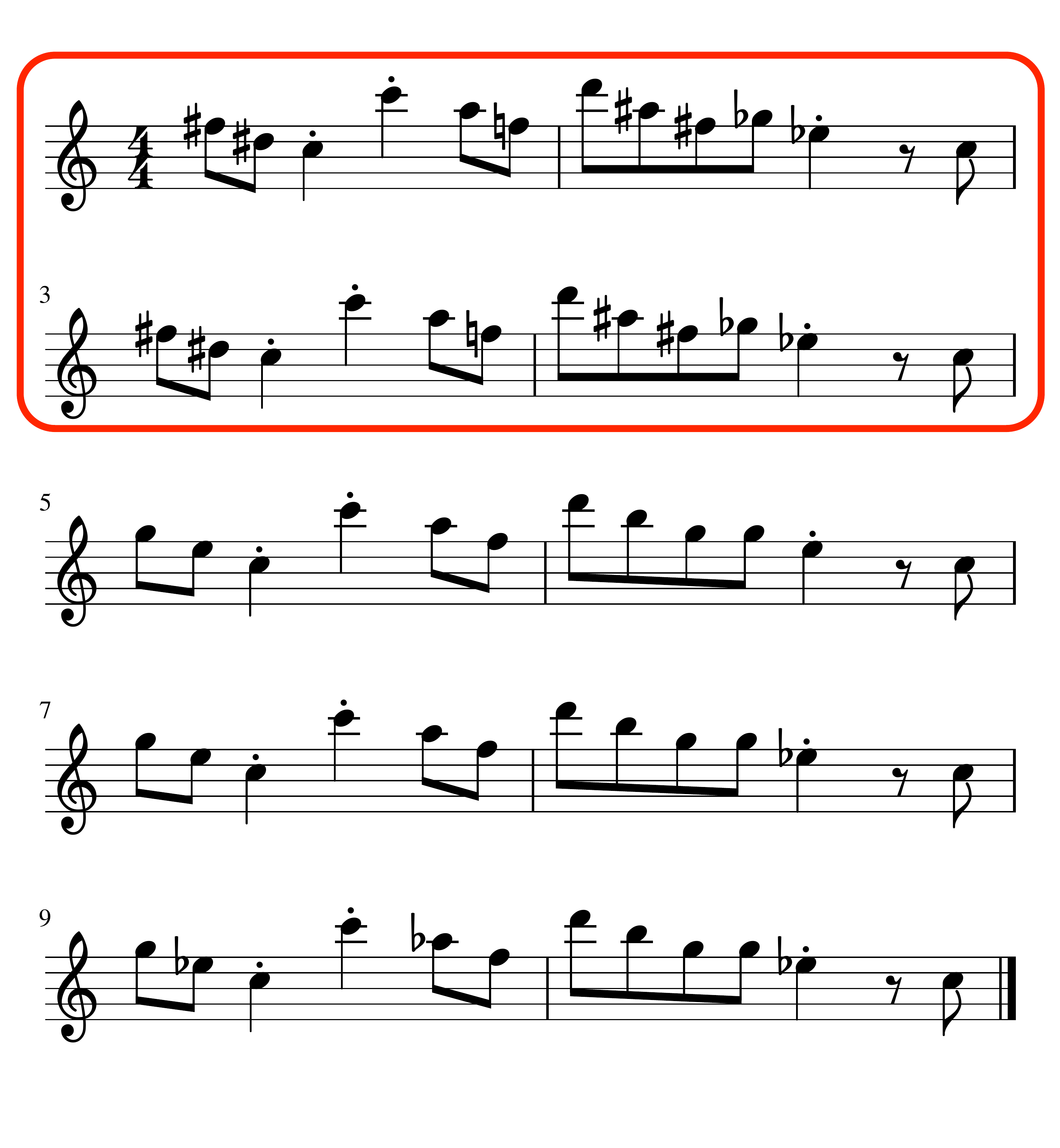}\vspace{-8pt} \\[\abovecaptionskip]
        \small (b) Scale
      \end{tabular}
      \begin{tabular}{@{}c@{}}
        \includegraphics[width=0.23\textwidth]{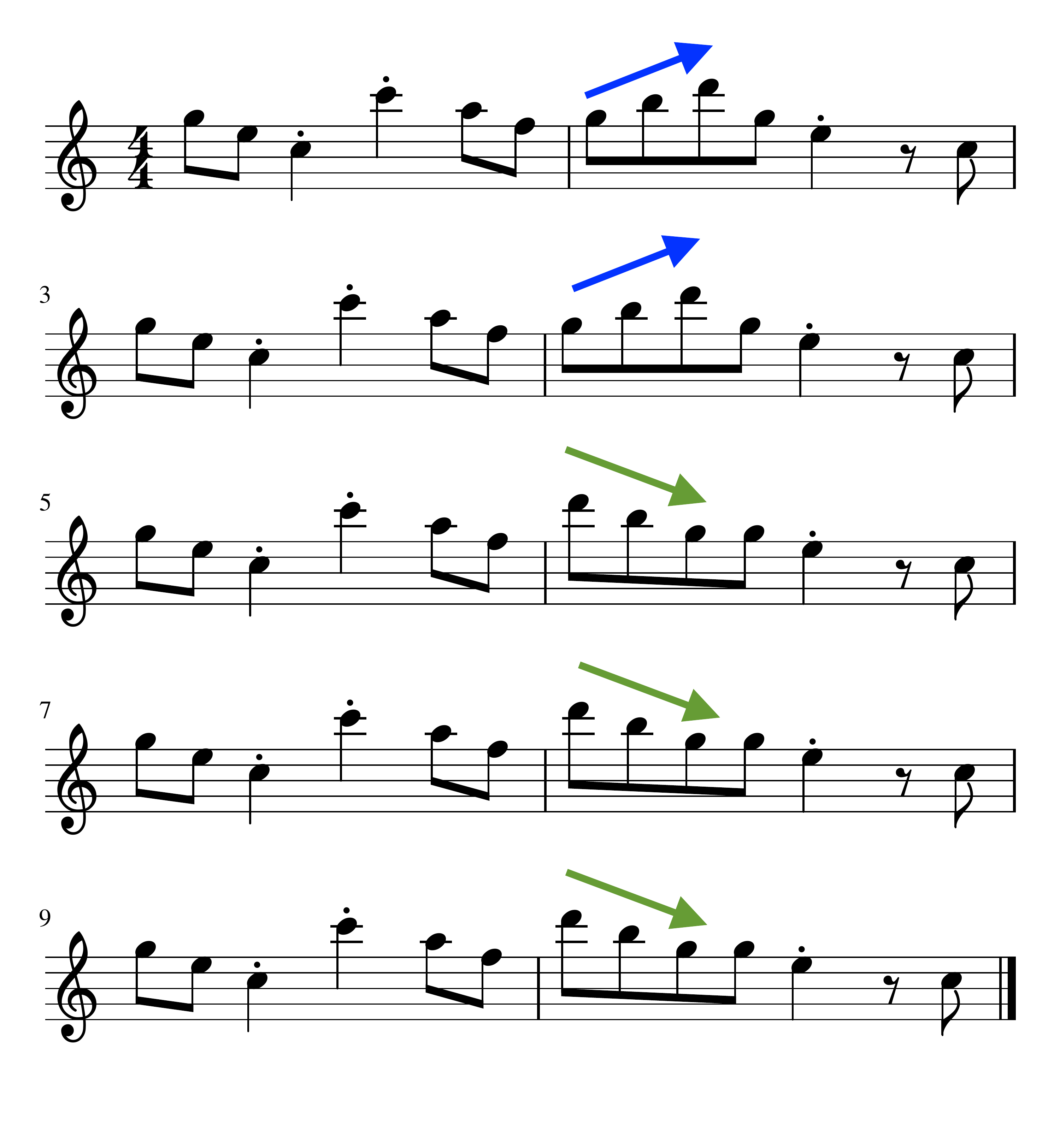}\vspace{-8pt} \\[\abovecaptionskip]
        \small (c) Arp Chord 3
      \end{tabular}
      \begin{tabular}{@{}c@{}}
        \includegraphics[width=0.23\textwidth]{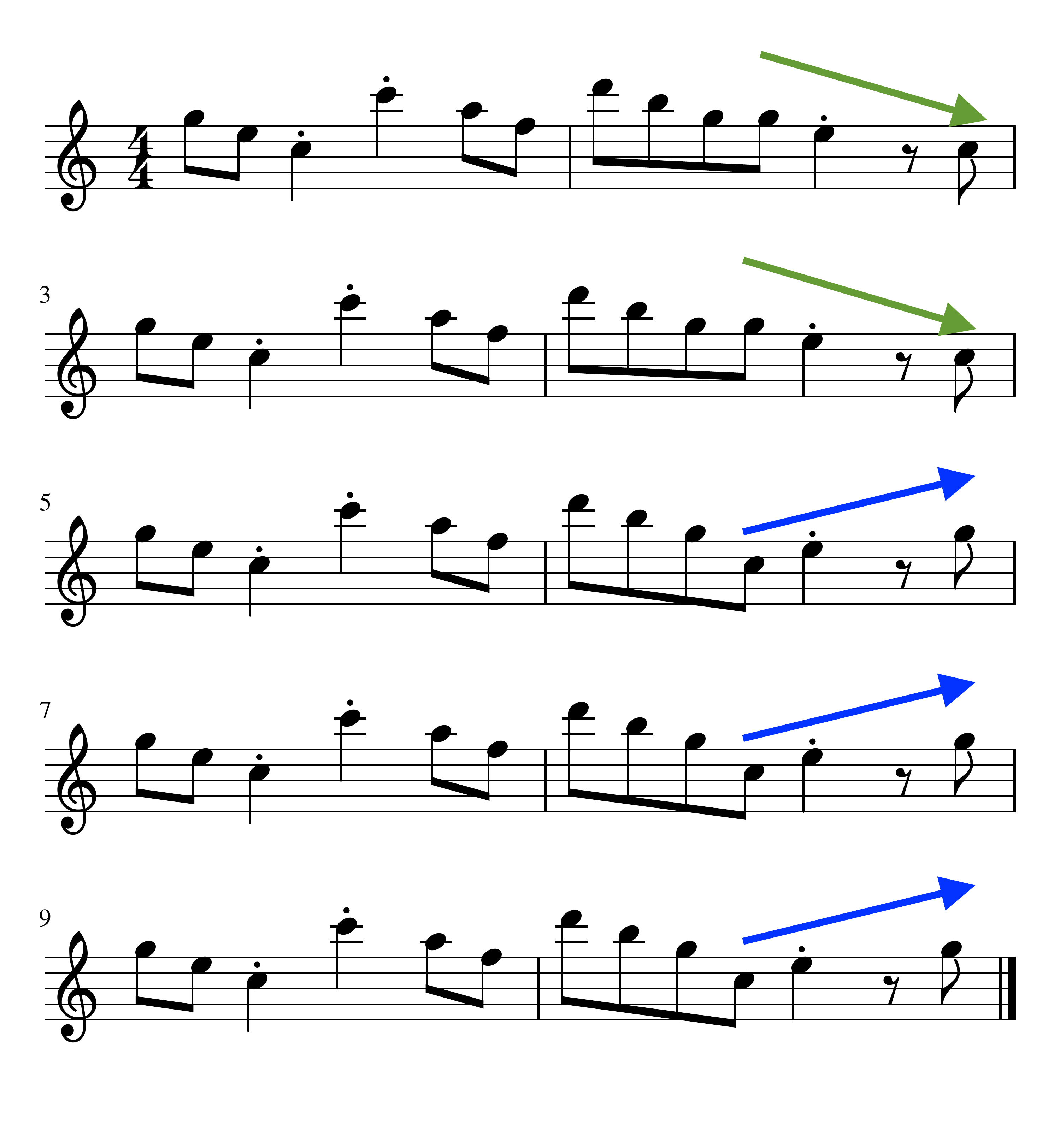}\vspace{-8pt} \\[\abovecaptionskip]
        \small (d) Arp Chord 4
      \end{tabular}
      \caption[Generated data by traversing along regularized dimensions of the I-VAE, dMelodies-RNN model]{Generated data by traversing along regularized dimensions for I-VAE.} 
    \label{fig:gen_data_IVAE}
    \end{figure*}

\begin{figure*}[t]
      \centering
      \begin{tabular}{@{}c@{}}
        \includegraphics[width=0.23\textwidth]{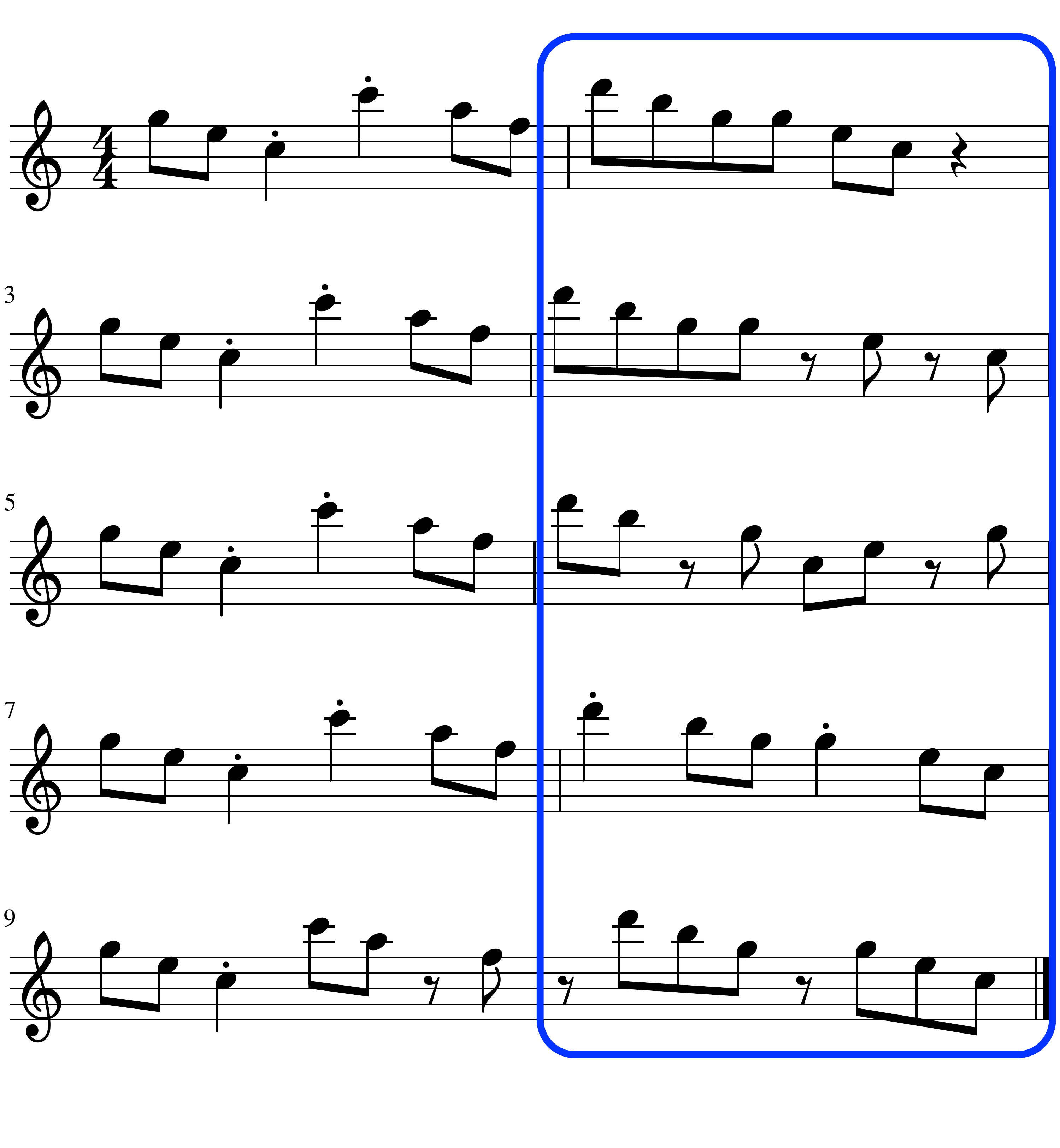}\vspace{-8pt} \\[\abovecaptionskip]
        \small (a) Rhythm Bar 2
      \end{tabular}
      \begin{tabular}{@{}c@{}}
        \includegraphics[width=0.23\textwidth]{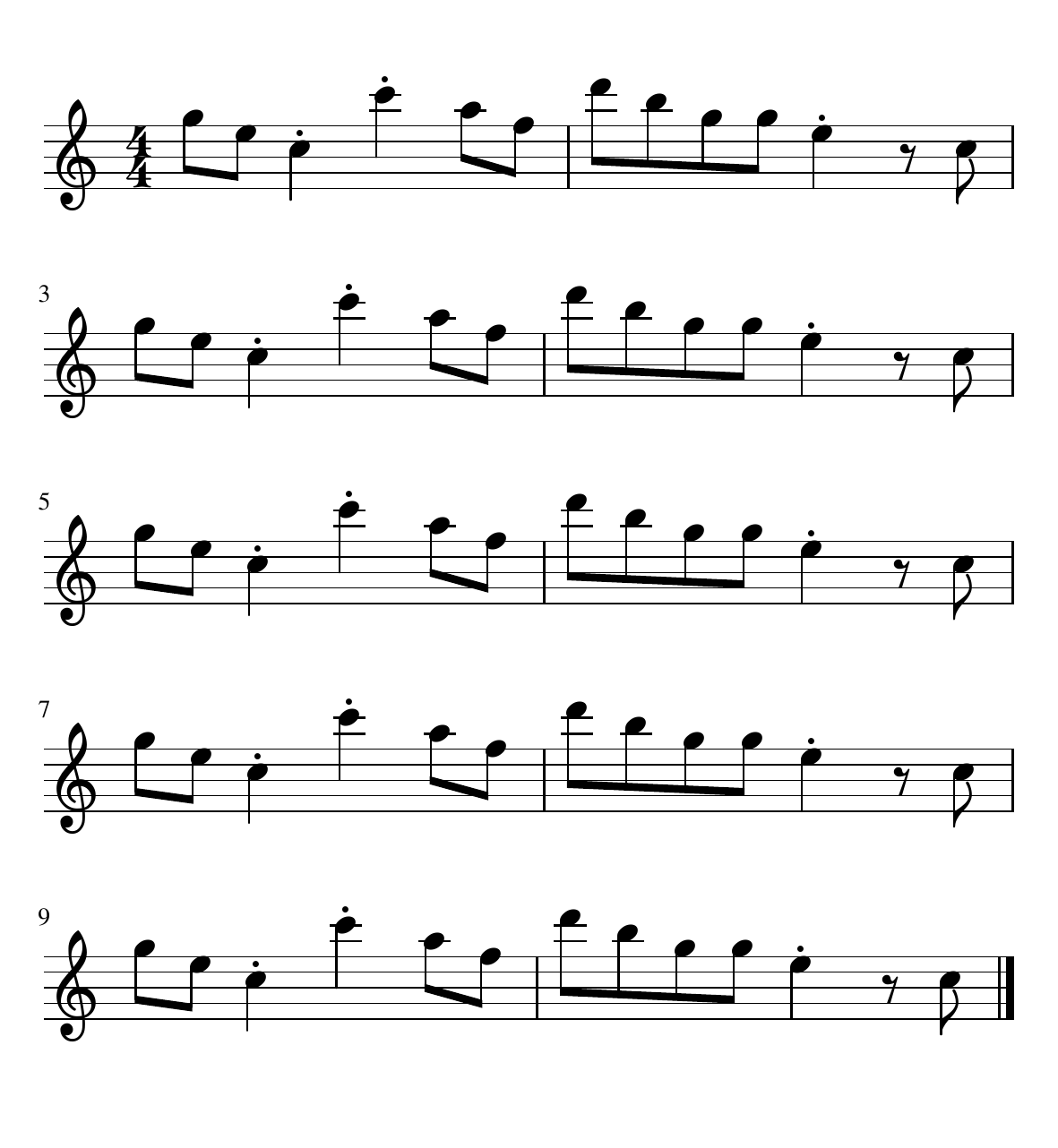}\vspace{-8pt} \\[\abovecaptionskip]
        \small (b) Scale
      \end{tabular}
      \begin{tabular}{@{}c@{}}
        \includegraphics[width=0.23\textwidth]{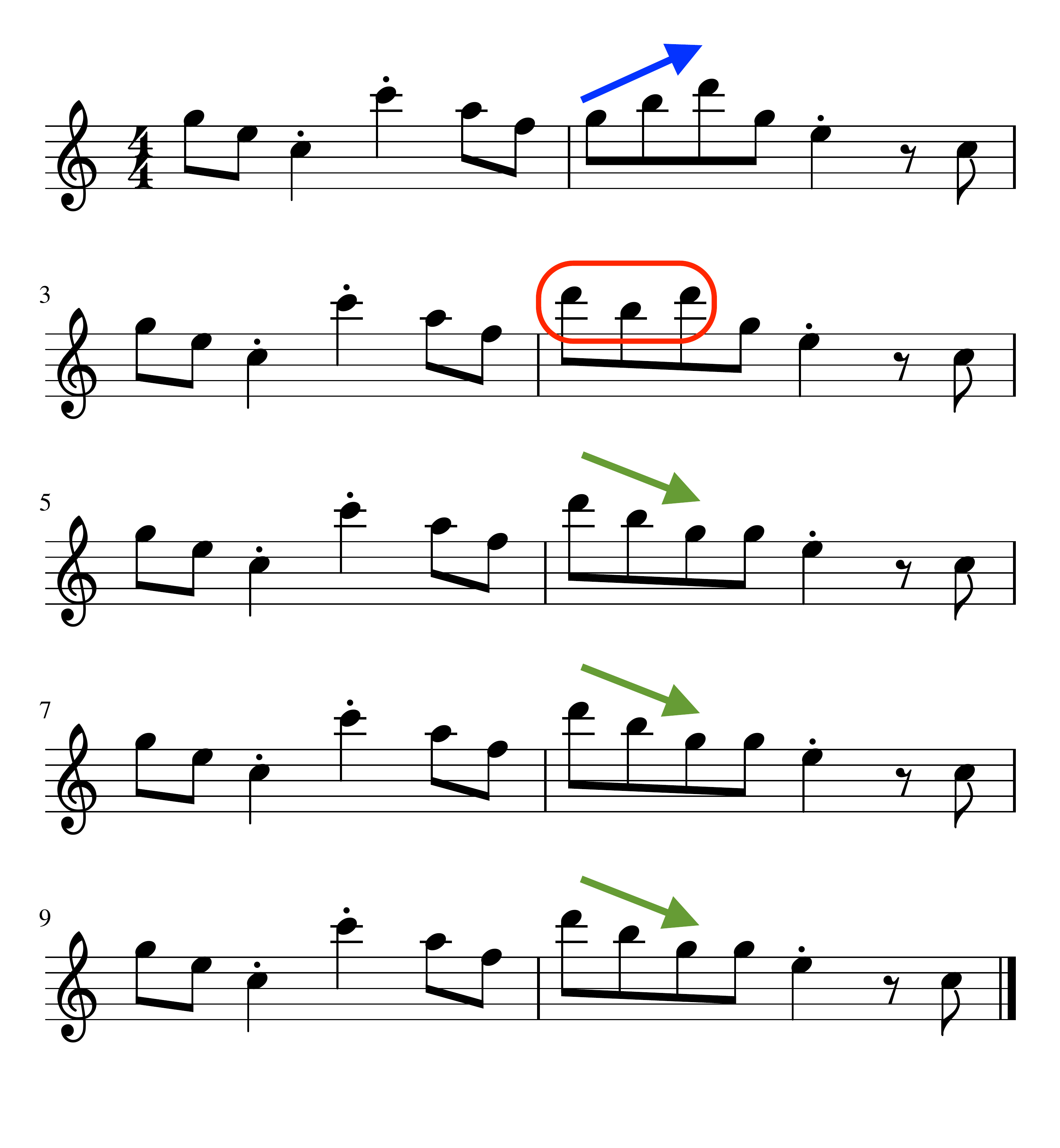}\vspace{-8pt} \\[\abovecaptionskip]
        \small (c) Arp Chord 3
      \end{tabular}
      \begin{tabular}{@{}c@{}}
        \includegraphics[width=0.23\textwidth]{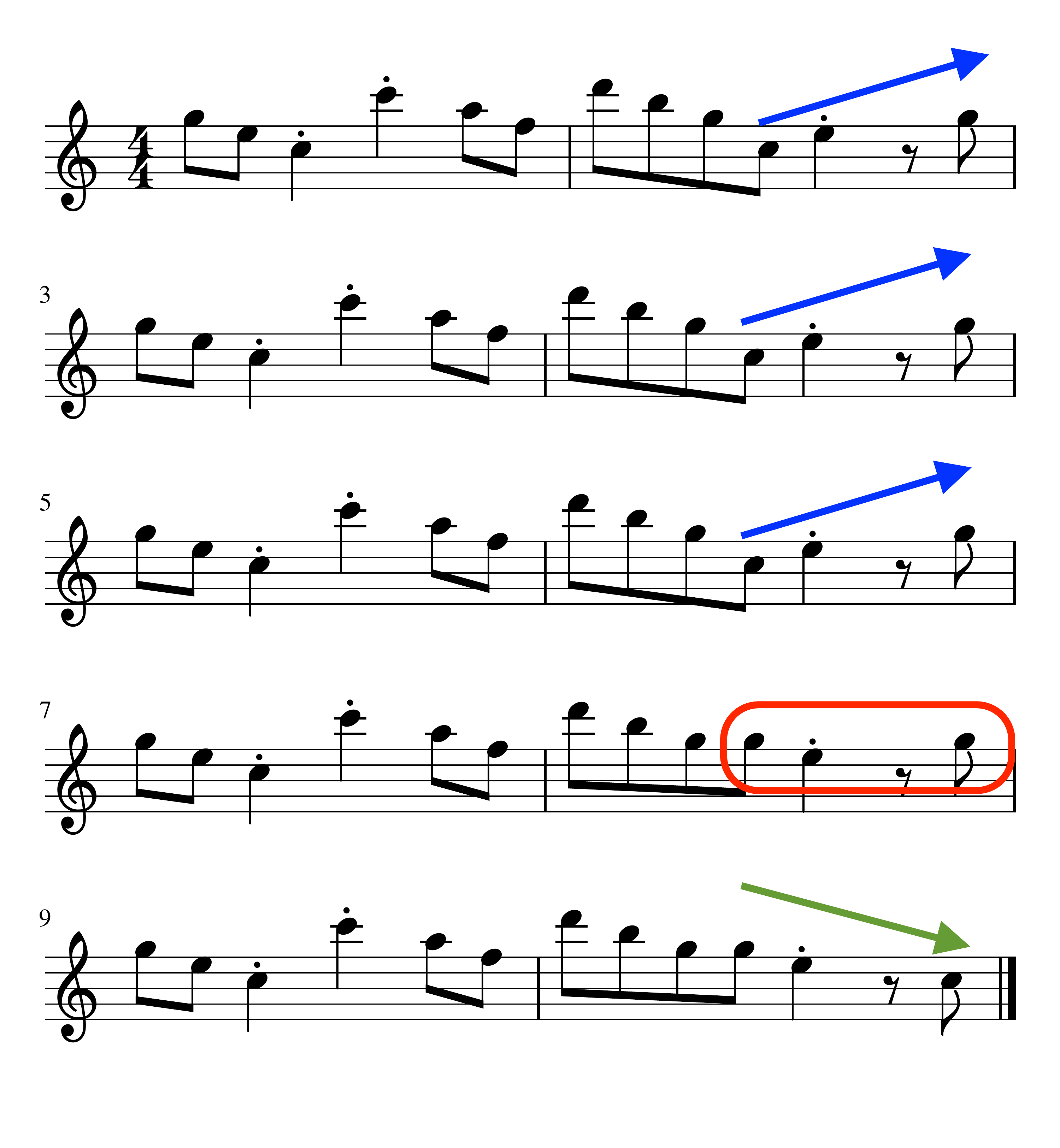}\vspace{-8pt} \\[\abovecaptionskip]
        \small (d) Arp Chord 4
      \end{tabular}
      \caption[Generated data by traversing along regularized dimensions of the AR-VAE, dMelodies-RNN model]{Generated data by traversing along regularized dimensions for AR-VAE.} 
    \label{fig:gen_data_arVAE}
    \end{figure*}

\begin{figure}[t]
      \centering
      \begin{tabular}{@{}c@{}}
        \includegraphics[width=0.23\textwidth]{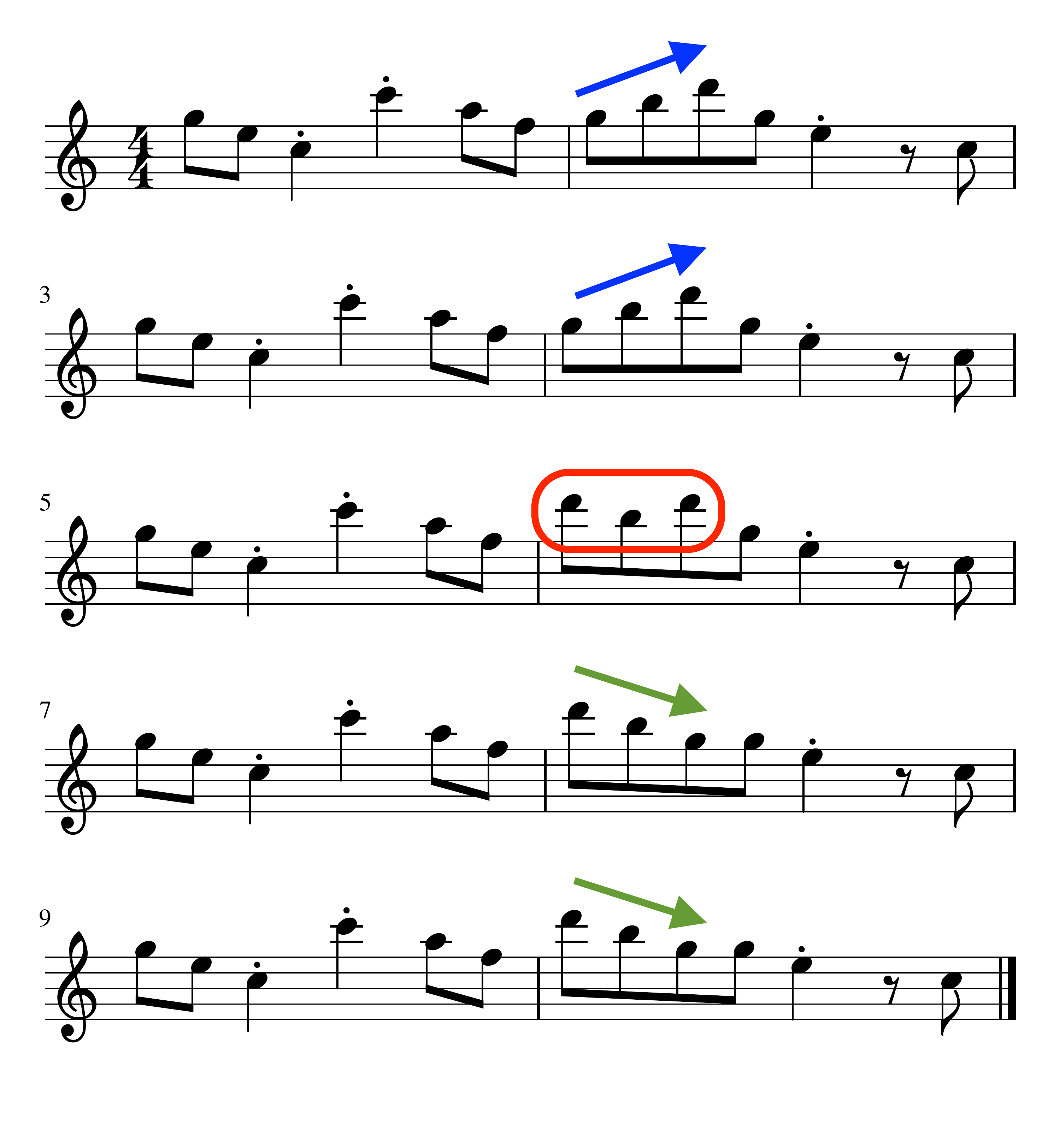}\vspace{-8pt} \\[\abovecaptionskip]
        \small (a) Arp Chord 3
      \end{tabular}
      \begin{tabular}{@{}c@{}}
        \includegraphics[width=0.23\textwidth]{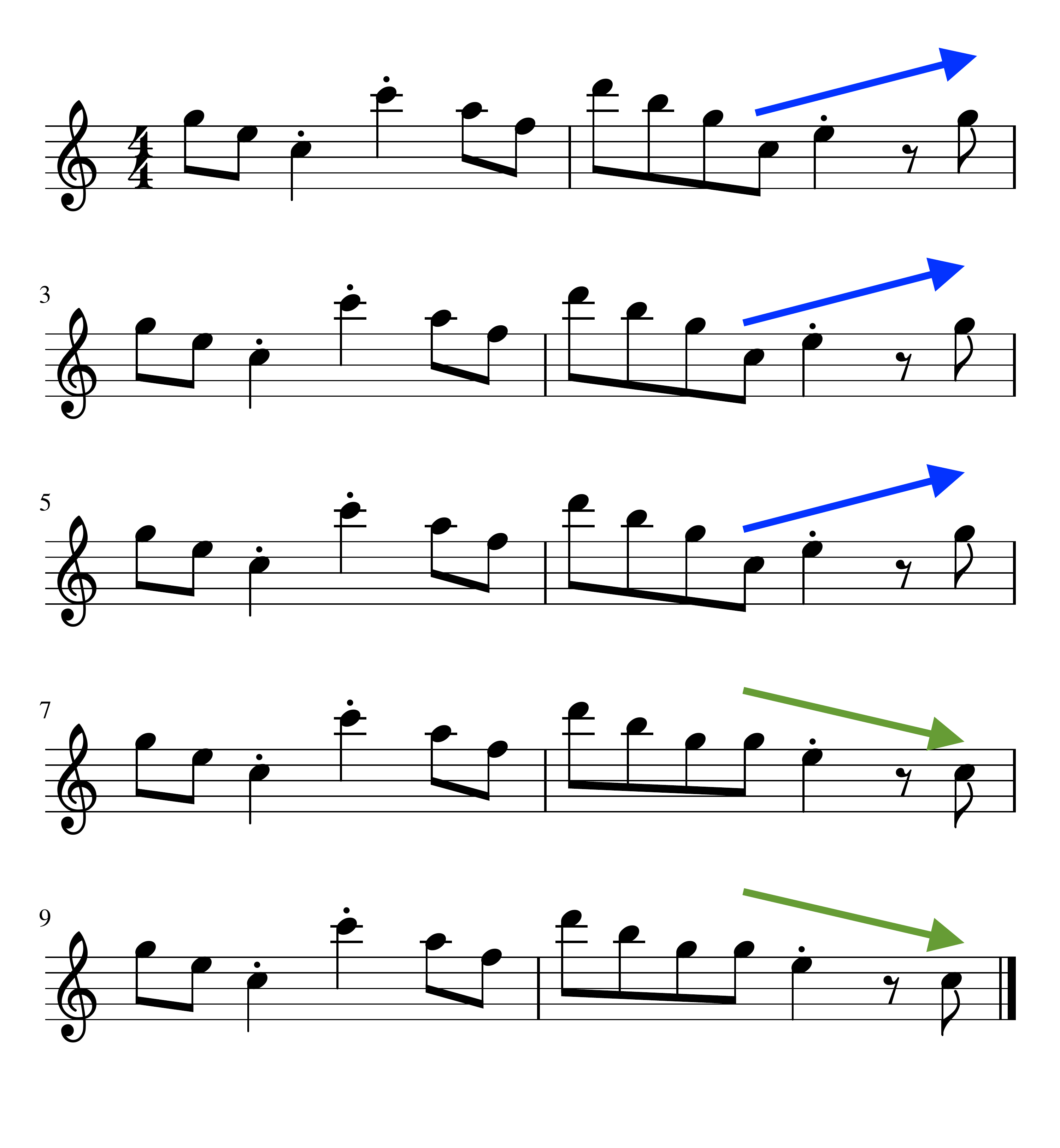}\vspace{-8pt} \\[\abovecaptionskip]
        \small (b) Arp Chord 4
      \end{tabular}
      \caption[Generated data by traversing along regularized dimensions of the S2-VAE, dMelodies-RNN model]{Generated data by traversing along regularized dimensions for S2-VAE.} 
    \label{fig:gen_data_s2VAE}
    \end{figure}

    Finally, we take a qualitative look at the data generated by the different methods while traversing the latent space along the regularized dimensions. Ideally, traversals along a regularized dimension should only cause changes in the corresponding attribute while leaving the other attributes unchanged. In addition, the regularized attribute should also change in a predictable manner. \figref{fig:gen_data_IVAE} shows the results for the I-VAE method. For each sub-figure, different rows correspond to melodies generated by traversing along the regularized dimension for the attribute in the sub-figure caption. Results for S2-VAE and AR-VAE are shown in Figures \ref{fig:gen_data_s2VAE} and \ref{fig:gen_data_arVAE}, respectively.

    \begin{table}[t]
      \footnotesize
      \begin{center}
          \begin{tabularx}{0.6\columnwidth}{Xlc}
              \toprule
              \textbf{Learning Method} & \textbf{LDR} \\ \toprule
              I-VAE & 0.448 \\ \midrule
              S2-VAE & 0.544 \\ \midrule
              AR-VAE & 0.244 \\ \bottomrule
          \end{tabularx}
      \end{center}
      \caption{LDR metric (higher is better) for different methods}
    \label{tab:ldm}
    \end{table}

    Across methods, most of the time, the melodies generated by traversing along regularized dimensions show changes in the corresponding attribute only. For instance, in Figures \ref{fig:gen_data_IVAE}(a) and \ref{fig:gen_data_arVAE}(a), only the rhythm of the second bar changes while the rest of the melody stays intact. In \figref{fig:gen_data_IVAE}(c,d), the arpeggiation directions of the third and fourth chords are flipped, respectively. Also, in \figref{fig:gen_data_IVAE}(b), all the other attributes remain constant (rhythm, arpeggiation directions) while the pitches of the generated notes change to reflect different scales. While this is desirable, there are a few important things to note. 

    First, the \textit{scale} attribute seems hard to control. For instance, in \figref{fig:gen_data_IVAE}(b), for I-VAE, some of the generated melodies (the first two rows) do not conform to any of the scales present in the dataset. In \figref{fig:gen_data_arVAE}(b), for AR-VAE, the \textit{scale} does not change at all. This difficulty in controlling the \textit{scale} attribute was also observed in \secref{dis:sup_attr_disent}. Second, depending on the holes in the latent space, traversals along regularized dimensions sometimes create melodies with attributes that are unseen in the training data. This happens also for attributes other than \textit{scale}. For instance, in \figref{fig:gen_data_arVAE}(c), row 2, the third chord has an unseen arpeggiation direction. Finally, for I-VAE, the direction of change for arpeggiation factors (see \figref{fig:gen_data_IVAE}(c,d)) is unpredictable. While the arpeggiation direction (of the third chord) goes from up to down in \figref{fig:gen_data_IVAE}(c), the direction (for the fourth chord) is flipped from down to up in \figref{fig:gen_data_IVAE}(d). This is due to the I-VAE regularization formulation which is agnostic to the order of the categorical attributes. Contrast this to AR-VAE and S2-VAE, where the nature of the change in the attribute values is predictable. The direction of arpeggiation will always go from up to down for these methods (see Figures \ref{fig:gen_data_s2VAE}(a,b) and \ref{fig:gen_data_arVAE}(c,d)).

  \subsection{Discussion}
    The results of the experiments in this section show that supervised methods for disentanglement perform significantly better than unsupervised methods. This is expected since the former use attribute-specific information during training to guide the model towards learning better representations. Among the supervised methods, there are no major differences in terms of the disentanglement metrics in \secref{dis:sup_disent}. However, controllability during data generation (discussed in Sections~\ref{dis:sup_attr_disent}, and \ref{dis:sup_latent_interp}) differs considerably between the methods. These differences suggest that while disentanglement is closely related to a strong encoder (learning the posterior $q(\mathbf{z} | \mathbf{x})$), improving controllability requires a strong decoder (learning the likelihood $p(\mathbf{x} | \mathbf{z})$). This explains the often better performance of conditioning-based methods relying on adversarial training of decoders \cite{lample_fader_2017, kawai2020attributes}. 
    
    Visualizing the latent spaces (in \secref{dis:sup_latent_viz}) with respect to the attribute values highlights that the presence or absence of holes in the learned latent space plays a crucial role in the degree of controllability afforded by a model. The LDR metric proposed in \secref{dis:ldm} is an attempt to quantify this behavior. Note that other factors can be considered while evaluating controllability that have been left out of this study. For instance, for continuous-valued attributes, one would prefer the regularized dimension having a positive correlation with the attribute value \cite{pati2020arvae}. 

\section{Conclusion} \label{sec:conclusion}
  In this paper, we present a systematic investigation of the relationship between attribute disentanglement and controllability in the context of symbolic music. Through a diverse set of experiments using different methods, we show that even though different supervised learning techniques can force effective disentanglement in the learned representations to a comparable extent, not all methods are equally effective at allowing control over the attributes during the data generation process. This distinction is important because controllability is paramount for generative models \cite{Briot2018} and is often not taken into account while evaluating disentanglement learning methods. 

  An important observation is the issue of holes in latent spaces. It should be noted this has also been seen in other data domains relying on discrete data such as text \cite{xu2019variational}. There are a few promising directions to address this problem. One option is to constrain the latent space to conform to a specific manifold and perform manipulations within this manifold \cite{xu2019variational, connor_representing_2020}. An alternative direction could be to learn specific transformation paths within the existing latent manifold to avoid these holes \cite{berthelot2018understanding}.

  The experiments in this paper have used labels from the entire training set. Another interesting direction for future studies could be to extend these experiments to a semi-supervised paradigm by using a limited number of labels obtained from only a fraction of the training set \cite{Locatello2020Disentangling}. This would increase the confidence in applying these methods to real-world data where obtaining label information for the entire dataset might be either too costly or simply impossible.

\section{Acknowledgments}
  The authors would like to thank NVIDIA Corporation (Santa Clara, CA, United States) for supporting this research via the NVIDIA GPU Grant program. 

\bibliography{ISMIRtemplate}

\begin{thebibliography}{10}
\providecommand{\url}[1]{#1}
\csname url@samestyle\endcsname
\providecommand{\newblock}{\relax}
\providecommand{\bibinfo}[2]{#2}
\providecommand{\BIBentrySTDinterwordspacing}{\spaceskip=0pt\relax}
\providecommand{\BIBentryALTinterwordstretchfactor}{4}
\providecommand{\BIBentryALTinterwordspacing}{\spaceskip=\fontdimen2\font plus
\BIBentryALTinterwordstretchfactor\fontdimen3\font minus
  \fontdimen4\font\relax}
\providecommand{\BIBforeignlanguage}[2]{{%
\expandafter\ifx\csname l@#1\endcsname\relax
\typeout{** WARNING: IEEEtran.bst: No hyphenation pattern has been}%
\typeout{** loaded for the language `#1'. Using the pattern for}%
\typeout{** the default language instead.}%
\else
\language=\csname l@#1\endcsname
\fi
#2}}
\providecommand{\BIBdecl}{\relax}
\BIBdecl

\bibitem{roberts_hierarchical_2018}
A.~Roberts, J.~Engel, C.~Raffel, C.~Hawthorne, and D.~Eck, ``A {Hierarchical}
  {Latent} {Vector} {Model} for {Learning} {Long}-{Term} {Structure} in
  {Music},'' in \emph{Proc. of 35th International {Conference} on {Machine}
  {Learning} ({ICML})}, Stockholm, Sweeden, 2018.

\bibitem{colombo2016algorithmic}
F.~Colombo, S.~Muscinelli, A.~Seeholzer, J.~Brea, and W.~Gerstner,
  ``{Algorithmic composition of melodies with deep recurrent neural
  networks},'' in \emph{Proc. of 1st Conference on Computer Simulation of
  Musical Creativity (CSMC)}, Huddersfield, UK, 2016.

\bibitem{sturm2016music}
B.~L. Sturm, J.~F. Santos, O.~Ben-Tal, and I.~Korshunova, ``{Music
  transcription modelling and composition using deep learning},'' in
  \emph{Proc. of 1st Conference on Computer Simulation of Musical Creativity
  (CSMC)}, Huddersfield, UK, 2016.

\bibitem{yang2017midinet}
L.-C. Yang, S.-Y. Chou, and Y.-H. Yang, ``{M}idi{N}et: A convolutional
  generative adversarial network for symbolic-domain music generation,'' in
  \emph{Proc. of 18th International Society of Music Information Retrieval
  Conference (ISMIR)}, Suzhou, China, 2017, pp. 324--331.

\bibitem{boulanger2012modeling}
N.~Boulanger-Lewandowski, Y.~Bengio, and P.~Vincent, ``{Modeling temporal
  dependencies in high-dimensional sequences: Application to polyphonic music
  generation and transcription},'' in \emph{Proc. of 29th International
  Conference on Machine Learning (ICML)}, Edinburgh, Scotland, 2012.

\bibitem{huang2019musictransformer}
C.-Z.~A. Huang, A.~Vaswani, J.~Uszkoreit, I.~Simon, C.~Hawthorne, N.~Shazeer,
  A.~M. Dai, M.~D. Hoffman, M.~Dinculescu, and D.~Eck, ``{Music transformer},''
  in \emph{Proc. of International Conference of Learning Representations
  (ICLR)}, New Orleans, USA, 2019.

\bibitem{oore2018time}
S.~Oore, I.~Simon, S.~Dieleman, D.~Eck, and K.~Simonyan, ``{This time with
  feeling: Learning expressive musical performance},'' \emph{Neural Computing
  and Applications}, pp. 1--13, 2018.

\bibitem{Briot2018}
J.-P. Briot and F.~Pachet, ``{Deep learning for music generation: Challenges
  and directions},'' \emph{Neural Computing and Applications}, 2018.

\bibitem{pati_learning_2019}
A.~Pati, A.~Lerch, and G.~Hadjeres, ``Learning to {Traverse} {Latent} {Spaces}
  for {Musical} {Score} {Inpainting},'' in \emph{Proc. of 20th International
  {Society} for {Music} {Information} {Retrieval} {Conference} ({ISMIR})},
  Delft, The Netherlands, 2019.

\bibitem{yang2019deep}
R.~Yang, D.~Wang, Z.~Wang, T.~Chen, J.~Jiang, and G.~Xia, ``Deep music analogy
  via latent representation disentanglement,'' in \emph{Proc. of 20th
  International {Society} for {Music} {Information} {Retrieval} {Conference}
  ({ISMIR})}, Delft, The Netherlands, 2019.

\bibitem{huang2019counterpoint}
C.-Z.~A. Huang, T.~Cooijmans, A.~Roberts, A.~Courville, and D.~Eck,
  ``Counterpoint by convolution,'' in \emph{{Proc. of the 18th International
  Society for Music Information Retrieval Conference (ISMIR)}}, Suzhou, China,
  2017.

\bibitem{hadjeres2017deepbach}
G.~Hadjeres, F.~Pachet, and F.~Nielsen, ``Deep{B}ach: {A} steerable model for
  {B}ach chorales generation,'' in \emph{Proc. of 34th International Conference
  on Machine Learning (ICML)}, Sydney, Australia, 2017, pp. 1362--1371.

\bibitem{donahue2019piano}
C.~Donahue, I.~Simon, and S.~Dieleman, ``Piano genie,'' in \emph{Proc. of 24th
  International Conference on Intelligent User Interfaces (IUI)}, Los Angeles,
  USA, 2019, pp. 160--164.

\bibitem{bazin2019nonoto}
T.~Bazin and G.~Hadjeres, ``Nonoto: A model-agnostic web interface for
  interactive music composition by inpainting,'' in \emph{Proc. of 10th
  International Conference on Computational Creativity (ICCC)}, UNC Charlotte,
  NC, USA, 2019.

\bibitem{bengio_representation_2013}
Y.~Bengio, A.~Courville, and P.~Vincent, ``Representation {Learning}: {A}
  {Review} and {New} {Perspectives},'' \emph{IEEE Transactions on Pattern
  Analysis and Machine Intelligence}, vol.~35, no.~8, 2013.

\bibitem{chen2020simple}
T.~Chen, S.~Kornblith, M.~Norouzi, and G.~Hinton, ``A simple framework for
  contrastive learning of visual representations,'' in \emph{Proc. of 37th
  International {Conference} on {Machine} {Learning} ({ICML})}.\hskip 1em plus
  0.5em minus 0.4em\relax PMLR, 2020, pp. 1597--1607.

\bibitem{caron2020unsupervised}
M.~Caron, I.~Misra, J.~Mairal, P.~Goyal, P.~Bojanowski, and A.~Joulin,
  ``Unsupervised learning of visual features by contrasting cluster
  assignments,'' in \emph{Advances in {Neural} {Information} {Processing}
  {Systems} 34 ({NeurIPS})}, 2020.

\bibitem{locatello_challenging_2019}
F.~Locatello, S.~Bauer, M.~Lucic, G.~Rätsch, S.~Gelly, B.~Schölkopf, and
  O.~Bachem, ``Challenging {Common} {Assumptions} in the {Unsupervised}
  {Learning} of {Disentangled} {Representations},'' in \emph{Proc. of 36th
  International {Conference} on {Machine} {Learning} ({ICML})}, Long Beach,
  California, USA, 2019.

\bibitem{jiang2020transformer}
J.~Jiang, G.~G. Xia, D.~B. Carlton, C.~N. Anderson, and R.~H. Miyakawa,
  ``{Transformer VAE: A Hierarchical Model for Structure-Aware and
  Interpretable Music Representation Learning},'' in \emph{Proc. of IEEE
  International Conference on Acoustics, Speech and Signal Processing
  (ICASSP)}, Barcelona, Spain, 2020, pp. 516--520.

\bibitem{brunner_midi-vae_2018}
G.~Brunner, A.~Konrad, Y.~Wang, and R.~Wattenhofer, ``{MIDI}-{VAE}: {Modeling}
  {Dynamics} and {Instrumentation} of {Music} with {Applications} to {Style}
  {Transfer},'' in \emph{Proc. of 19th International {Society} for {Music}
  {Information} {Retrieval} {Conference} ({ISMIR})}, Paris, France, 2018.

\bibitem{hung2019musical}
Y.-N. Hung, I.-T. Chiang, Y.-A. Chen, and Y.-H. Yang, ``Musical composition
  style transfer via disentangled timbre representations,'' in \emph{Proc. of
  28th International Joint Conference on Artificial Intelligence (IJCAI)},
  Macao, China, 2020.

\bibitem{luo2019learning}
Y.-J. Luo, K.~Agres, and D.~Herremans, ``Learning disentangled representations
  of timbre and pitch for musical instrument sounds using gaussian mixture
  variational autoencoders,'' in \emph{Proc. of 20th International {Society}
  for {Music} {Information} {Retrieval} {Conference} ({ISMIR})}, Delft, The
  Netherlands, 2019.

\bibitem{hadjeres_glsr-vae_2017}
G.~Hadjeres, F.~Nielsen, and F.~Pachet, ``{GLSR}-{VAE}: {Geodesic} latent space
  regularization for variational autoencoder architectures,'' in \emph{Proc. of
  {IEEE} {Symposium} {Series} on {Computational} {Intelligence} ({SSCI})},
  Hawaii, USA, 2017, pp. 1--7.

\bibitem{pati19latent-reg}
A.~Pati and A.~Lerch, ``Latent space regularization for explicit control of
  musical attributes,'' in \emph{Proc. of ICML Workshop on Machine Learning for
  Music Discovery Workshop (ML4MD), Extended Abstract}, Long Beach, California,
  USA, 2019.

\bibitem{tan2020music}
H.~H. Tan and D.~Herremans, ``{Music FaderNets: Controllable music generation
  based on high-level features via low-level feature modelling},'' in
  \emph{Proc. of 20th International Society for Music Information Retrieval
  Conference (ISMIR)}, Montréal, Canada, 2020.

\bibitem{Locatello2020Disentangling}
F.~Locatello, M.~Tschannen, S.~Bauer, G.~Rätsch, B.~Schölkopf, and O.~Bachem,
  ``Disentangling factors of variations using few labels,'' in \emph{Proc. of
  8th International Conference on Learning Representations (ICLR)}, Addis
  Ababa, Ethiopia, 2020.

\bibitem{pati2020dmelodies}
A.~Pati, S.~Gururani, and A.~Lerch, ``{dMelodies: A Music Dataset for
  Disentanglement Learning},'' in \emph{Proc. of 21st International Society for
  Music Information Retrieval Conference (ISMIR)}, Montréal, Canada, 2020.

\bibitem{kingma_auto-encoding_2014}
D.~P. Kingma and M.~Welling, ``Auto-{Encoding} {Variational} {Bayes},'' in
  \emph{Proc. of 2nd International {Conference} on {Learning} {Representations}
  ({ICLR})}, Banff, Canada, 2014.

\bibitem{adel_discovering_2018}
T.~Adel, Z.~Ghahramani, and A.~Weller, ``Discovering {Interpretable}
  {Representations} for {Both} {Deep} {Generative} and {Discriminative}
  {Models},'' in \emph{Proc. of 35th International {Conference} on {Machine}
  {Learning} ({ICML})}, Stockholm, Sweden, 2018, pp. 50--59.

\bibitem{pati2020arvae}
\BIBentryALTinterwordspacing
A.~Pati and A.~Lerch, ``{Attribute-based Regularization of Latent Spaces for
  Variational Auto-Encoders},'' \emph{Neural Computing and Applications}, 2020.
  [Online]. Available: \url{https://doi.org/10.1007/s00521-020-05270-2}
\BIBentrySTDinterwordspacing

\bibitem{higgins_beta-vae_2017}
I.~Higgins, L.~Matthey, A.~Pal, C.~Burgess, X.~Glorot, M.~M. Botvinick,
  S.~Mohamed, and A.~Lerchner, ``beta-{VAE}: {Learning} {Basic} {Visual}
  {Concepts} with a {Constrained} {Variational} {Framework},'' in \emph{Proc.
  of 5th International {Conference} on {Learning} {Representations} ({ICLR})},
  Toulon, France, 2017.

\bibitem{kingma_adam_2015}
D.~P. Kingma and J.~Ba, ``Adam: {A} {Method} for {Stochastic} {Optimization},''
  in \emph{Proc. of 3rd International {Conference} on {Learning}
  {Representations} ({ICLR})}, San Diego, USA, 2015.

\bibitem{chen_isolating_2018}
R.~T.~Q. Chen, X.~Li, R.~Grosse, and D.~Duvenaud, ``Isolating {Sources} of
  {Disentanglement} in {Variational} {Autoencoders},'' in \emph{Advances in
  {Neural} {Information} {Processing} {Systems} 32 ({NeurIPS})}, Montréal,
  Canada, 2018.

\bibitem{ridgeway_learning_2018}
K.~Ridgeway and M.~C. Mozer, ``Learning {Deep} {Disentangled} {Embeddings}
  {With} the {F}-{Statistic} {Loss},'' in \emph{Advances in {Neural}
  {Information} {Processing} {Systems} 32 ({NeurIPS})}, Montréal, Canada,
  2018, pp. 185--194.

\bibitem{kumar_variational_2017}
A.~Kumar, P.~Sattigeri, and A.~Balakrishnan, ``Variational {Inference} of
  {Disentangled} {Latent} {Concepts} from {Unlabeled} {Observations},'' in
  \emph{Proc. of 5th International {Conference} of {Learning} {Representations}
  ({ICLR})}, Toulon, France, 2017.

\bibitem{lample_fader_2017}
G.~Lample, N.~Zeghidour, N.~Usunier, A.~Bordes, L.~Denoyer, and M.~Ranzato,
  ``Fader {Networks}:{Manipulating} {Images} by {Sliding} {Attributes},'' in
  \emph{Advances in {Neural} {Information} {Processing} {Systems} 31
  ({NeurIPS})}, Long Beach, California, USA, 2017, pp. 5967--5976.

\bibitem{kawai2020attributes}
L.~Kawai, P.~Esling, and T.~Harada, ``Attributes-aware deep music
  transformation,'' in \emph{Proc. of 21st International {Society} for {Music}
  {Information} {Retrieval} {Conference} ({ISMIR})}, Montréal, Canada, 2020.

\bibitem{xu2019variational}
P.~Xu, J.~C.~K. Cheung, and Y.~Cao, ``On variational learning of controllable
  representations for text without supervision,'' in \emph{Proc. of 37th
  International Conference on Machine Learning (ICML)}, 2020.

\bibitem{connor_representing_2020}
M.~Connor and C.~Rozell, ``Representing {Closed} {Transformation} {Paths} in
  {Encoded} {Network} {Latent} {Space},'' in \emph{Proc. of 34th {AAAI}
  {Conference} on {Artificial} {Intelligence}}, New York, USA, 2020.

\bibitem{berthelot2018understanding}
D.~Berthelot, C.~Raffel, A.~Roy, and I.~Goodfellow, ``Understanding and
  improving interpolation in autoencoders via an adversarial regularizer,'' in
  \emph{Proc. of 7th International {Conference} on {Learning} {Representations}
  ({ICLR})}, New Orleans, USA, 2019.

\end{thebibliography}
\end{document}